\documentclass[twocolumn,showpacs,preprintnumbers,amsmath,amssymb]{revtex4}

\usepackage{graphicx}
\usepackage{dcolumn}
\usepackage{bm}
\usepackage{mathrsfs}
\usepackage{amsmath}

\begin{document}

\preprint{APS/123-QED}

\title{Finite-size scaling of string order parameters characterizing the Haldane phase}

\author{Hiroshi Ueda}
\email{ueda@aquarius.mp.es.osaka-u.ac.jp}
\author{Hiroki Nakano$^1$}
\email{hnakano@sci.u-hyogo.ac.jp}
\author{Koichi Kusakabe}
\email{kabe@mp.es.osaka-u.ac.jp}

\affiliation{Graduate School of Engineering Science, Osaka University, 1-3 Machikaneyama-cho, Toyonaka, Osaka 560-8531, Japan}
\affiliation{$^1$Graduate School of Material Science, University of Hyogo, 3-2-1 Kouto, Kamigori-cho, Akou-gun, 678-1297, Japan}

\date{\today}

\begin{abstract}
We have developed a numerical procedure to clarify 
the critical behavior near a quantum phase transition 
by analyzing a multi-point correlation function 
characterizing the ground state. 
This work presents a successful application 
of this procedure to the string order parameter 
of the $S=1$ $XXZ$ chain with uniaxial single-ion anisotropy. 
The finite-size string correlation function is estimated 
by the density matrix renormalization group method. 
We focus on the gradient of the inversed-system-size dependence 
of the correlation function on a logarithmic plot. 
This quantity shows that the finite-size scaling 
sensitively changes at the critical point. 
The behavior of the gradient with increasing system size 
is divergent, stable at a finite value, 
or rapidly decreases to zero 
when the system is in the disordered phase, 
at the critical point, or in the ordered phase, respectively. 
The analysis of the finite-size string correlation functions 
allows precise determination of the boundary 
of the Haldane phase and 
estimation of the critical exponent of the correlation length.
Our estimates of the transition point and the critical exponents, 
which are determined only by the ground-state quantities,
are consistent with results obtained from the analysis 
of the energy-level structure. 
Our analysis requires only the correlation functions of several 
finite sizes under the same condition as a candidate for the long-range 
order. 
The quantity is treated in the same manner irrespective of 
the kind of elements which destroy the order concerned.
This work will assist in the development of a method to directly observe 
quantum phase transitions. 
\end{abstract}

\maketitle

\section{\label{sec:level1}Introduction}
Quantum phase transitions originating 
from quantum fluctuations have been extensively studied 
as a hot issue in condensed-matter physics. 
Several interesting characteristics of the transitions appear 
in the low-energy behavior of the systems. 
Two types of approach 
can capture the phase transitions and critical phenomena 
precisely when the transition is continuous.  
One is analyzing the energy-level structure. 
The other involves considering 
the ground-state behavior. 

In the former approach, a standard method is 
to analyze the structure of the energy levels 
of finite-size systems 
based on the finite-size scaling (FSS) assumption. 
For example, the scaled energy gap \cite{Barber:FSS} is often 
used to estimate the boundary of the gapped phase 
as the transition point. 
This is called phenomenological renormalization group (PRG) 
analysis. 
However, it is difficult to estimate the transition point 
when a logarithmic correction appears 
in the dependence of the energy difference. 
A typical example is the Berezinskii--Kosterlitz--Thouless (BKT) 
Transition \cite{Kosterlitz:JPC7}. 
To resolve this difficulty, the level-spectroscopy method 
has been developed \cite{Nomura:JPA28} 
and precise determinations of phase transitions have been successfully made 
for various transitions in many models. 
Unfortunately, this analysis is complicated 
in that 
appropriate adjustments of the procedure are required 
according to the type of phase transition, 
which must also be known in advance.  

In the latter approach, on the other hand, 
quantities 
that characterize the ground state are carefully observed. 
One of these quantities is the multi-point correlation function. 
The long-range behavior of correlation functions shows 
whether the system exhibits long-range order. 
If a correlation function survives to be nonzero 
in the long-range limit,  
it is an appropriate order parameter. 
However, it is not easy to capture a phase transition 
using this strategy 
because reliable and precise data on correlation functions 
are necessary for large systems. 
The system sizes that are treated 
in numerical-diagonalization calculations are insufficient. 
For this reason, 
the latter approach has been employed in only a few studies. 
Therefore, 
no systematic procedure for analyzing ground-state quantities 
to capture quantum phase transitions 
has been established to date. 

In this paper, we develop a procedure to determine 
the transition point and critical exponents 
by analyzing correlation functions 
based only on the scaling assumption. 
A feature of this approach is that 
only the common quantities under the same condition 
are treated irrespective of the type of phase transition. 
We call the procedure ground-state 
phenomenological renormalization group (GSPRG) analysis. 
To confirm its validity and usefulness 
in detecting phase transitions, 
we apply it to a nontrivial ground state in 
the AF $S=1$ $XXZ$ chain with uniaxial single-ion anisotropy 
by the density matrix renormalization group (DMRG) 
method \cite{White:PRL69, White:PRB48}. 

In the isotropic case of this system, 
there exists a nonzero energy gap 
between the unique ground state and the first excited state, 
called 
the Haldane gap\cite{Haldane:PL93A, Haldane:PRL50}. 
It is known that when anisotropy is introduced, 
of the single-ion type or of the $XXZ$-type exchange interaction, 
the Haldane gap decreases and finally closes. 
The region where the nonzero Haldane gap exists 
is called the Haldane phase. 
The phase diagram of the AF $S=1$ chain 
with anisotropy of the two types, including the Haldane phase, 
has been extensively studied  by analyzing 
the energy-level structure, 
assisted by the level-spectroscopy method \cite{Alcaraz:PRB46, Hida:PRB67, 
Boschi:EPJB35, Boschi:EPJB41}.  

It is well known that 
in many AF spin systems, 
the standard spin--spin correlation function 
gives so-called N$\acute{\textrm{e}}$el order. 
In the ground state in the Haldane phase, however, 
the spin--spin correlation function decays exponentially 
with a finite correlation length and 
the N$\acute{\textrm{e}}$el order no longer exists. 
In this sense, the Haldane phase is a disordered phase. 
However, 
the string order parameter is 
known to characterize the ground state in the Haldane phase, 
in which the longitudinal and transverse string order 
parameters are nonzero \cite{Nijs:PRB40, Kennedy:PRB45}. 
From the viewpoint of the string order, 
it is possible to treat the Haldane phase 
as an ordered phase and capture the phase transition 
at the boundary. 
This approach has been applied 
to numerical-diagonalization data 
of the Haldane phase 
in $S=1$ systems\cite{Alcaraz:PRB46, Totsuka:JPCM7, 
Tonegawa:JPSJ65}. 
Unfortunately, only very small system sizes were available 
and hence it was quite difficult to capture precisely
the critical behavior of the string order 
near the transition point.  

In this situation, 
we can obtain numerical data of the string order 
of this model for much larger system sizes by using DMRG for 
GSPRG analysis. 
Consequently, it is possible to examine the phase transition 
at the boundary of the Haldane phase 
from the viewpoint of the string order. 
We compare our results with those from the analysis 
of the energy-level structure. 
This comparison 
provides a systematic and consistent understanding 
of the phase transition. 

This paper is organized as follows. 
The model Hamiltonian and order parameters are defined 
in section \ref{sec:level2}. 
The analysis procedure which we have developed 
is introduced in section \ref{sec:level3}. 
The numerical results and discussions are given 
in section \ref{sec:level4}.
Section \ref{sec:level5} consists of a summary of this work and 
some remarks.

\section{\label{sec:level2}Model Hamiltonian and order parameters}
We consider the following Hamiltonian:
\begin{eqnarray}
\mathcal{H} \! = \!\! \sum^{N-1}_{i=1} \!
\left[
J \! \left(S^x_i S^x_{i+1} \! + \! S^y_i S^y_{i+1} \right) \! + \! J_z S^z_i S^z_{i+1}
\right]
\!\! + \! D \! \sum^N_{i=1} \! {S^z_i}^2 ,
\label{eq.1}
\end{eqnarray}
where $N$ is the system size, $S^\alpha_i (\alpha = x,y,z)$ are spin-1 operators, 
$J$ and $J_z$ represent the $XXZ$-type anisotropic exchange interaction, 
and the parameter $D$ represents uniaxial single-ion anisotropy. 
In this paper, energies are measured in units of $J$ 
and hence we take $J=1$ hereafter.  
The boundary condition of the system is open.
An antiferromagnetic chain is usually characterized 
by the N$\acute{\textrm{e}}$el order parameter defined as
\begin{equation}
O^\alpha_{\textrm{N}\acute{\textrm{e}}\textrm{el}}  = \lim_{|i-j| \rightarrow \infty} 
O^\alpha_{\textrm{N}\acute{\textrm{e}}\textrm{el}}(i,j),
\label{eq.2b} 
\end{equation}
where $O^\alpha_{\textrm{N}\acute{\textrm{e}}\textrm{el}}(i,j)$ 
is the N$\acute{\textrm{e}}$el correlation function given by
\begin{equation}
O^\alpha_{\textrm{N}\acute{\textrm{e}}\textrm{el}}(i,j)  = (-1)^{i-j} \langle S^\alpha_i S^\alpha_j \rangle. 
\label{eq.2a}
\end{equation}
Here $\langle \hat{B} \rangle$ represents 
the ground-state expectation value 
of an arbitrary operator $\hat{B}$.
In the Haldane phase, 
the N$\acute{\textrm{e}}$el order parameter vanishes.
However,  
the string order introduced 
by den Nijs and Rommelse \cite{Nijs:PRB40} appears instead. 
The string order parameter is given by
\begin{equation}
O^\alpha_{\textrm{str}}  = \lim_{|i-j| \rightarrow \infty} O^\alpha_{\textrm{str}}(i,j),
\label{ep.3b}
\end{equation}
where the string correlation function $O^\alpha_{\textrm{str}}(i,j)$ 
is given by
\begin{equation}
O^\alpha_{\textrm{str}}(i,j)  = - \left\langle S^\alpha_i \exp \left[ i\pi \sum^{j-i}_{k=i+1} S^\alpha_k \right] S^\alpha_j \right\rangle.
\label{ep.3a} 
\end{equation}

Kennedy and Tasaki \cite{Kennedy:PRB45} extensively studied 
the string order and applied a nonlocal unitary transformation 
to the Hamiltonian (\ref{eq.1}). 
Thereby they obtained
\begin{align}
O^\alpha_{\textrm{str}} & = O^\alpha_{\textrm{ferro}}(\tilde{\mathcal{H}}) \:\:\: \textrm{for} \: \alpha=x\:\textrm{or}\:z,
\label{ep.4}
\end{align}
where 
$O^\alpha_{\textrm{ferro}} 
= \lim_{|i-j| \rightarrow \infty}
\left\langle S^\alpha_i S^\alpha_j \right\rangle$, 
and $\tilde{\mathcal{H}}$ is obtained by applying the transformation to 
the original Hamiltonian $\mathcal{H}$.  
In the transformed system, 
$\mathbb{Z}_2\times\mathbb{Z}_2$ symmetry emerges. 
Breaking of this symmetry is described 
by the behavior of the order parameters 
$O^\alpha_{\textrm{ferro}}(\tilde{\mathcal{H}})$. 
When the system is in the Haldane phase, 
this $\mathbb{Z}_2\times\mathbb{Z}_2$ symmetry is fully broken. 
When the chain is in a phase other than the Haldane phase, 
the full symmetry or a part of the symmetry survives 
\cite{Kennedy:PRB45, Alcaraz:PRB46, Hatsugai:PRB44}. 

\section{\label{sec:level3}Analysis procedure}
In this section we introduce our analysis procedure. 
The procedure consists of three steps. 
The first is to calculate 
the longitudinal and transverse order parameters 
by the DMRG method based 
on the finite-size algorithm \cite{White:PRL69, White:PRB48}. 
Knowing the behavior of the order parameters enables us 
to observe the boundary of the Haldane phase briefly and 
to confirm that the transition at the boundary is continuous. 
If we apply the FSS analysis of the order parameters \cite{Tonooka:JPSJ76}, 
we can roughly estimate the critical point and exponents. 
However, there remain finite-size effects in these estimates. 
In order to eliminate the effects, we carry out 
a finite-size extrapolation derived 
from the FSS formula. 
As the second step, we introduce a finite-size quantity 
which we calculate from the string order parameters. 
This quantity reaches the critical exponent just at the phase boundary 
when we take the limit $N\rightarrow\infty$. 
By examining the behavior of this quantity, 
it is possible to obtain the phase boundary 
and the critical exponent of the order parameter at the transition point. 
At the third step, 
we estimate the critical exponent of the correlation length 
near the transition point by extrapolating a finite-size 
quantity at the above transition point.  
The present procedure 
gives consistent and 
precise estimates for the critical point and exponents. 

\subsection{\label{sec:level3-1}Calculation of order parameters}
To calculate the order parameters, we use the finite DMRG algorithm 
with the acceleration algorithm introduced 
by White \cite{White:PRL69, White:PRB48, White:PRL77}. 
A correlation function, such 
as Eq. (\ref{eq.2a}) or Eq. (\ref{ep.3a}), is calculated 
as follows. 
First, 
we obtain a variational wave function 
of the ground state represented by the matrix product state (MPS): 
\begin{align}
\left| \Phi \right\rangle  = \hspace{0mm} \sum_{\{s\}} \sum_{\{\alpha\}} \hspace{0mm} 
\left( U_{s_1, \alpha_2}^{[2], s_2} U_{\alpha_2, \alpha_3}^{[3], s_3} \cdots 
U_{\alpha_{N/2-2}, \alpha_{N/2-1}}^{[N/2-1], s_{N/2-1}} \right. \hspace{10mm} \nonumber \\
 A_{\alpha_{N/2-1}, \alpha_{N/2+2}}^{s_{N/2}, s_{N/2+1}} 
V_{\alpha_{N/2+2}, \alpha_{N/2+3}}^{[N/2+2], s_{N/2+2}} \cdots \hspace{9mm} \nonumber \\
 \left. V_{\alpha_{N-2}, \alpha_{N-1}}^{[N-2], s_{N-2}} V_{\alpha_{N-1}, s_N}^{[N-1], s_{N-1}} \right)
\left| s_1 s_2 \cdots s_N\right\rangle , \nonumber \\
(\{s\}=s_1, s_2, \cdots, s_N, \hspace{2mm} \{\alpha\}=\alpha_2, \alpha_3, \cdots, \alpha_{N-1} ), 
\label{eq.5-5}
\end{align}
where $U$ and $V$ are matrices satisfying 
\begin{subequations}
\begin{align}
U^{[i]\dagger} U^{[i]} &= \mathbf{1}, &(i=2, \cdots, N/2-1), 
\label{eq.5-7a}\\
V^{[j]} V^{[j] \dagger} &= \mathbf{1}, &(j=N/2+2, \cdots, N-1),
\label{eq.5-7b}
\end{align}
\end{subequations}
respectively, and 
$A$ is the ground state of the renormalized Hamiltonian which 
is obtained by applying the density matrix renormalization 
transformation to the Hamiltonian (\ref{eq.1}).
A truncation error $\epsilon$ due to the cut off $m_{\rm F}$ is estimated 
by $1 - \textrm{Tr}\left( A^{*} A \right)$, where $m_{\rm F}$ is the number of states preserved in the DMRG iterations \cite{White:PRL69}.
Using this wave function, 
we estimate the expectation value 
$O^\alpha_{\textrm{str}}(i, j)$ as follows: 
\begin{align}
O^\alpha_{\textrm{str}}(i, j)  =  - \left\langle \Phi \right| S^\alpha_{i} \tilde{S}^\alpha_{i+1} \cdots 
\tilde{S}^\alpha_{j-1} S^\alpha_{j} \left| \Phi \right\rangle \nonumber \\ 
= - \textrm{Tr} \left( E^{[2]}_{\mathbf{1}, \mathbf{1}} E^{[3]}_{\mathbf{1}} \cdots E^{[i-1]}_{\mathbf{1}} 
E^{[i]}_{S^\alpha_{i}} E^{[i+1]}_{\tilde{S}^\alpha_{i+1}} \right. \cdots \hspace{8mm} \nonumber\\
E^{[N/2-1]}_{\tilde{S}^\alpha_{N/2-1}} 
E^{[A]}_{\tilde{S}^\alpha_{N/2}, \tilde{S}^\alpha_{N/2+1}}  
E^{[N/2+2]}_{\tilde{S}^\alpha_{N/2+2}} \cdots \hspace{5mm} \nonumber \\
\left. 
E^{[j-1]}_{\tilde{S}^\alpha_{j-1}} E^{[j]}_{S^\alpha_{j}} E^{[j+1]}_{\mathbf{1}} \cdots 
E^{[N-2]}_{\mathbf{1}} E^{[N-1]}_{\mathbf{1}, \mathbf{1}} 
\right). 
\label{eq.5-8}
\end{align}
Here, $\mathbf{1}$ is the identity matrix in spin-1 space 
and $\tilde{S}^\alpha$ is defined as 
$\tilde{S}^\alpha = \exp (i \pi S^\alpha)$. 
Each $E$ is a matrix for an arbitrary local operator $O$ defined by 
\begin{subequations}
\begin{align}
E^{[2]}_{O, O} \! & = & \sum_{s_1,s_1',s_2, s_2'}
\left\langle s_1 \right| O \left| s_1' \right\rangle \left\langle s_2 \right| O \left| s_2' \right\rangle \hspace{23mm} \nonumber \\
&&U_{s_1}^{[2], s_2} \otimes \left( U_{s_1'}^{[2], s_2'} \right)^*, \hspace{4mm}
\label{eq.5-9a}\\
E^{[i]}_{O} & = & 
\sum_{s_i,s_i'} \left\langle s_i \right| O \left| s_i' \right\rangle  U_{}^{[i], s_i} 
\otimes \left( U_{}^{[i], s_i'} \right)^*, \hspace{14mm} \nonumber \\
&&(3 \leq i \leq N/2-1), \hspace{5mm} 
\label{eq.5-9b} \\
E^{[A]}_{O, O} & = & \sum_{s_k,s_k', s_{k+1},s_{k+1}'} \!\!\!\!\!\!
\left\langle s_k \right| O \left| s_k' \right\rangle \left\langle s_{k+1} \right| O \left| s_{k+1}' \right\rangle   \nonumber \hspace{13mm} \\
&&A_{}^{s_k, s_{k+1}} \otimes 
\left( A_{}^{s_k', s_{k+1}'} \right)^*, \nonumber \hspace{5mm} \\
&&(k=N/2), \hspace{5mm} 
\label{eq.5-9c}\\
E^{[j]}_{O} & = & 
\sum_{s_j,s_j'} \left\langle s_j \right| O \left| s_j' \right\rangle  V_{}^{[j], s_j} 
\otimes \left( V_{}^{[j], s_{j}'} \right)^*, \hspace{10mm} \nonumber \\
&&(N/2+2 \leq j \leq N-2), \hspace{5mm} 
\label{eq.5-9d}\\
E^{[N-1]}_{O, O} & = & 
\sum_{s_{N-1},s_{N-1}', s_N, s_N'} \hspace{-8mm}
\left\langle s_{N-1} \right| O \left| s_{N-1}' \right\rangle  \left\langle s_N \right| O \left| s_N' \right\rangle  \nonumber   \hspace{11mm} \\
&&V_{s_N}^{[N-1], s_{N-1}} 
\otimes \left( V_{s_N'}^{[N-1], s_{N-1}'} \right)^*. \hspace{5mm} 
\label{eq.5-9e}
\end{align}
\end{subequations}
When calculating a two-point correlation function 
such as Eq. (\ref{eq.2a}), 
we replace all the operators $\tilde{S}^\alpha$ with $\mathbf{1}$. 
We also choose an appropriate correlation function 
as the longest-ranged component 
from among $O^\alpha_{\textrm{str}}(i, j)$ or 
$O^\alpha_{\textrm{N}\acute{\textrm{e}}\textrm{el}}(i, j)$ 
for a fixed $N$ under the open boundary condition. 
Three desirable conditions should be met, as follows.
\begin{enumerate}
\item The measurement points $i$ and $j$ are as far as possible 
from the edges. 
\item The correlation distance of $|j - i|$ is as long 
as possible. 
\item The distance $|j - i|$ should increase in proportion to the system size $N$. 
\end{enumerate}
In order to satisfy the above conditions, 
we take $i=N/3+1$ and $j=2N/3$. 
Thus, we consider the order parameter 
\begin{align}
O^\alpha(N, D, J_z) & = O^\alpha(N/3+1, 2N/3, D, J_z), 
\label{eq.5}
\end{align}
where $\alpha = x, z$, $O^\alpha$ represents 
$O^\alpha_{\textrm{N}\acute{\textrm{e}}\textrm{el}}$ or 
$O^\alpha_{\textrm{str}}$. 
Note here that $N/6$ should be an integer. 
We emphasize again that Eq. (\ref{eq.5}) is useful 
in the case of the open boundary condition \cite{comment_PBC}. 

The error of the order parameters is estimated as follows. 
\begin{enumerate}
\item $\left\langle  O^\alpha \right\rangle_{m_1}$ is calculated 
in the case of $m_1 = 0.95 \times m_{\textrm{F}}$. 
\item $\left\langle  O \right\rangle_{m=m_2}$ and 
the truncation error $\epsilon$ are calculated 
in the case of $m_2 = m_{\textrm{F}}$. 
\item The error estimation of the order parameters $\left\langle  O^\alpha \right\rangle^{\textrm{error}}_{m_{\textrm{F}}}$ is defined by 
the following formula: 
\begin{align}
\left\langle  O^\alpha \right\rangle^{\textrm{error}}_{m_{\textrm{F}}} = 
\max \left[ \left| \left\langle  O^\alpha \right\rangle_{m_2} - \left\langle  O^\alpha \right\rangle_{m_1} \right|, 
\left| \epsilon \left\langle  O^\alpha \right\rangle_{m_2} \right| \right].
\label{eq.6}
\end{align}
\end{enumerate}
In this paper, all numerical data have a truncation error $\epsilon < 10^{-7}$.

\subsection{\label{sec:level3-2}Finite-size scaling analysis}
In the general theory of phase transitions, 
the treatment of physical quantities depends on 
whether the transition is continuous or discontinuous. 
If the transition is continuous, 
the critical behavior of bulk quantities 
is extracted through the FSS analysis of finite-size quantities \cite{Barber:FSS}. 
As we observe later, the DMRG data of correlation functions 
of the finite-size systems are continuous 
near the boundary of the Haldane phase. 
The exact-diagonalization data 
of the string order parameters are also continuous. 
Thus, it is possible to perform FSS analysis 
of our DMRG data of the string order parameter. 

The present model includes two control parameters, 
$D$ and $J_z$. 
When $J_z$ is fixed and $D$ is varied, 
we carry out the FSS analysis based 
on the following equation: 
\begin{align}
O^\alpha_{\textrm{str}}(N, D, J_z^{\textrm{fix}}) = N^{-\eta_\alpha}\Psi((D_{\textrm{c}}-D)N^{1/\nu_\alpha}),
\label{eq.7}
\end{align}
where $\alpha = x$ or $z$, $J_z^{\textrm{fix}}$ is 
the fixed $J_z$, and $D_{\textrm{c}}$ is the critical point. 
The same equation concerning the string-type order parameter 
was used in Ref.~\cite{Tonooka:JPSJ76}. 
The exponents $\eta_\alpha$ and $\nu_\alpha$ are 
defined as 
\begin{subequations}
\begin{align}
\xi(N=\infty, D, J_z^{\textrm{fix}}) & \sim (D_{\textrm{c}} - D)^{-\nu},
\label{eq.8a} \\
O^\alpha_{\textrm{str}}(N, D_{\textrm{c}}, J_z^{\textrm{fix}}) & \sim N^{-\eta_\alpha},
\label{eq.8b}
\end{align}
\end{subequations}
where $\xi(N, D, J_z)$ represents the correlation length. 
When $D$ is fixed and $J_z$ is varied, 
on the other hand, the FSS formula is given by 
\begin{align}
O^\alpha_{\textrm{str}}(N, D^{\textrm{fix}}, J_z) = N^{-\eta_\alpha}\Psi((J^{\textrm{c}}_z-J_z)N^{1/\nu_\alpha}),
\label{eq.9}
\end{align}
where the critical exponents $\eta_\alpha$ and $\nu_\alpha$ are given by 
\begin{subequations}
\begin{align}
\xi(N=\infty, D^{\textrm{fix}}, J_z) & 
\sim (J^{\textrm{c}}_z - J)^{-\nu},
\label{eq.10a} \\
O^\alpha_{\textrm{str}}(N, D^{\textrm{fix}}, J^{\textrm{c}}_z) & \sim N^{-\eta_\alpha}.
\label{eq.10b}
\end{align}
\end{subequations}
Note that as 
the BKT transition point is approached, 
only Eq. (\ref{eq.8b}) or Eq. (\ref{eq.10b}) is realized. 
In this case, the correlation length grows 
exponentially \cite{Kosterlitz:JPC7}; 
the dependence is different from Eq. (\ref{eq.8a}) or Eq. (\ref{eq.10a}), 
having a finite exponent. 

Successfully obtaining a universal function $\Psi$ 
irrespective of the system size $N$ in the critical region near the critical point, 
allows us to determine the critical point and critical exponents. 
However, the FSS analysis still has a problem 
in that the width of the critical region is unknown. 
Since the width depends on the values of the control parameters, 
it is difficult to determine or estimate an appropriate width. 
To avoid this difficulty, 
we perform the extrapolation explained below. 

\subsection{\label{sec:level3-3}Ground-state 
phenomenological-renormalization-group analysis}
In this paper, we perform a procedure to obtain 
the transition point and 
the critical exponents consistently 
considering the ground-state quantities. 
We call this procedure the ground-state 
phenomenological-renormalization-group (GSPRG) analysis. 
The first step of the procedure 
is to examine a finite-size quantity defined as
\begin{align}
\eta_\alpha(\tilde{N}, D, J_z) \!=\! 
\frac{ \log \left[ O^\alpha_{\textrm{str}}(N_2 , D, J_z) 
/ O^\alpha_{\textrm{str}}(N_1, D, J_z) \right] }
{ \log[N_1/N_2] }, 
\label{eq.11}
\end{align}
where $\tilde{N}=(N_1+N_2)/2$, $N_2 = \Delta N + N_1$, and $\Delta N = 6$. 
Here, we examine the direction $\alpha$ such that 
$O^\alpha_{\textrm{str}}$ shows critical behavior.   
The quantity indicates the gradient of the curve 
of the dependence of $O^\alpha_{\textrm{str}}$ 
on $1/N$ in a plot with both the axes on the logarithmic scale. 
The gradient should be constant for large system sizes 
when the set of $D$ and $J_z$ corresponds to the boundary 
of the Haldane phase.  
The quantity $\eta_\alpha(\tilde{N}, D, J_z)$ 
converges to the critical exponent $\eta_{\alpha}$ 
defined by Eq. (\ref{eq.8b}) or Eq. (\ref{eq.10b}) 
for ($D$, $J_z$) on the boundary;  
the $\tilde{N}$ dependence of $\eta_\alpha(\tilde{N}, D, J_z)$ 
shows a stable convergence to a finite value 
when the system size is increased.  
On the other hand, when the point ($D$, $J_z$) is not on the boundary, 
$\eta_\alpha(\tilde{N}, D, J_z)$ shows 
a different behavior. 
For ($D$, $J_z$) inside the Haldane phase, 
$O^\alpha_{\textrm{str}}$ tends towards a nonzero value as $N$ is increased. 
Thus the gradient $\eta_\alpha(\tilde{N}, D, J_z)$ rapidly decreases.  
For ($D$, $J_z$) outside the Haldane phase, 
$O^\alpha_{\textrm{str}}$ decays rapidly with increasing $N$. 
This decay is more rapid than 
that for Eq. (\ref{eq.8b}) or Eq. (\ref{eq.10b}).  
In this case, the gradient $\eta_\alpha(\tilde{N}, D, J_z)$ 
rapidly increases. 
Therefore, 
we can find the critical point from the difference 
in the $\tilde{N}$-dependence of $\eta_\alpha(\tilde{N}, D, J_z)$.  
The difference is expected to be more apparent 
when $\tilde{N}$ increases sufficiently to diminish 
the edge effect. 
In order to estimate the critical point, we have investigated the behavior 
of $\eta_\alpha(\tilde{N}, D, J_z)$  
for $N=6,12,\cdots,90,96$ and found the characteristic behavior 
of $\eta_\alpha(\tilde{N}, D, J_z)$  
in the region of large $\tilde{N}$. 
The numerical procedure to determine the critical point 
by observing the behavior of $\eta_\alpha(\tilde{N}, D, J_z)$  
is as follows. 
\begin{itemize}
\item When the differentiation of the finite-size quantity  
$\eta_\alpha(\tilde{N}, D, J_z)$ satisfies 
the following conditions for a large system size,
\begin{subequations}
\begin{align}
\eta^{(1)}_\alpha(\tilde{N}, D, J_z) > 0 \wedge \eta^{(2)}_\alpha(\tilde{N}, D, J_z) > 0, \label{eq.12a} \\
{\rm or} \hspace{3cm} \nonumber \\
\eta^{(1)}_\alpha(\tilde{N}, D, J_z) < 0 \wedge \eta^{(2)}_\alpha(\tilde{N}, D, J_z) < 0, \label{eq.12b}
\end{align}
\end{subequations}
we can consider that 
the system with $(D, J_z)$ is at a critical point. 
Here $\eta^{(n)}_\alpha(\tilde{N}, D, J_z)(n=1,2)$ is 
the numerical differentiation given by
\begin{align}
\eta^{(n)}_\alpha (\tilde{N}, D, J_z) = 
\left( \frac{\partial}{\partial(1/\tilde{N})}  \right)^n \eta_\alpha(\tilde{N}, D, J_z).
\label{eq.13}
\end{align}
The differentiation is approximated by the difference because 
$\tilde{N}$ is integer or half integer.
\item If the differentiation reveals 
\begin{align}
\eta^{(1)}_\alpha(\tilde{N}, D, J_z) > 0 \wedge \eta^{(2)}_\alpha(\tilde{N}, D, J_z) < 0, \label{eq.15}
\end{align}
we can consider that 
$\eta_\alpha(\tilde{N}, D, J_z)$ will decrease rapidly 
with increasing system size. 
In this case,
the system with $(D, J_z)$ is in the ordered phase 
with respect to the string order.
\item If the differentiation satisfies
\begin{align}
\eta^{(1)}_\alpha(\tilde{N}, D, J_z) < 0 \wedge \eta^{(2)}_\alpha(\tilde{N}, D, J_z) > 0, \label{eq.14}
\end{align}
we can consider that 
$\eta_\alpha(\tilde{N}, D, J_z)$ will increase rapidly 
with increasing system size. 
In this case,
the system with $(D, J_z)$ is in the disordered phase 
with respect to the string order.
\end{itemize}
We summarize the difference in the behavior 
of $\eta_\alpha(\tilde{N}, D, J_z)$ in Table \ref{tab.2}. 
Now, the boundary for a finite-size system 
between the critical region and the string-ordered phase 
is given by $N^{\prime}$ defined in 
\begin{subequations}
\begin{align}
\eta^{(1)}_\alpha(N', D, J_z) > 0 \wedge \eta^{(2)}_\alpha(N', D, J_z) = 0, 
\label{co-boundary1} \\
{\rm or} \hspace{3cm} \nonumber \\
\eta^{(1)}_\alpha(N', D, J_z) = 0 \wedge \eta^{(2)}_\alpha(N', D, J_z) < 0, 
\label{co-boundary2}
\end{align}
\end{subequations}
to find a boundary 
between Eq. (\ref{eq.12a}, \ref{eq.12b}) and Eq. (\ref{eq.15}).
We obtain $N'$ as a real positive number 
because $\eta^{(i)}_\alpha(N', D, J_z)$  is 
an interpolated value of 
$\eta^{(i)}_\alpha(\tilde{N}, D, J_z)$ for $i=1,2$. 
The boundary between the critical region and 
the string disordered phase is, on the other hand, given 
by $N^{\prime}$ defined in 
\begin{subequations}
\begin{align}
\eta^{(1)}_\alpha(N', D, J_z) = 0 \wedge \eta^{(2)}_\alpha(N', D, J_z) > 0, 
\label{cd-boundary1} \\
{\rm or} \hspace{3cm} \nonumber \\
\eta^{(1)}_\alpha(N', D, J_z) < 0 \wedge \eta^{(2)}_\alpha(N', D, J_z) = 0, 
\label{cd-boundary2}
\end{align}
\end{subequations}
to find a boundary between Eq. (\ref{eq.12a}, \ref{eq.12b}) and Eq. (\ref{eq.14}). 
When the critical behavior appears only at a point, 
the width between the critical-ordered boundary and 
the critical-disordered boundary shrinks as $N'$ increases. 
Such behavior will be presented in section \ref{sec:level4-2}. 
In this work, we consider the width of the critical region 
between the two boundaries to be an error in our analysis 
of the transition point if the width is very narrow. 
In a case of the BKT transition, 
the critical-disordered boundary does not appear. 
In this case, we must estimate the transition point carefully 
only from the critical-ordered boundary.  
Details of this treatment will be given 
in section \ref{sec:level4-4}. 
\begin{table}
\caption{\label{tab.2}
Behavior of $\eta_\alpha(\tilde{N},D,J_z)$ 
in the four phases and near the three transition lines.
HN, HL, HX: 
Haldane-N$\acute{\textrm{e}}$el, 
Haldane-Large-$D$, Haldane-$XY$. 
RD, RI, SF:
rapidly decreasing finite-size exponent (\ref{eq.11}), 
rapidly increasing exponent, 
stably finite exponent, as the system size is increased. 
}
\begin{ruledtabular}
\begin{tabular}{cccc}
       & Haldane & HN transition line & N$\acute{\textrm{e}}$el \\
\hline
$\eta_x(\tilde{N}, D, J_z)$ & RD & SF & RI \\
$\eta_z(\tilde{N}, D, J_z)$ & RD & RD  & RD \\
\hline
       & Haldane & HL transition line & Large-$D$ \\
\hline
$\eta_\alpha(\tilde{N}, D, J_z)$ & RD & SF & RI \\
\hline
       & Haldane & HX transition line & $XY$ \\
\hline
$\eta_\alpha(\tilde{N}, D, J_z)$ & RD & SF & SF \\
\end{tabular}
\end{ruledtabular}
\end{table}

The estimation of $\eta_{\alpha}$ of the string order parameter 
by PRG analysis 
has been reported by Hida \cite{Hida:JPSJ62} 
based on finite-size data of the string orders 
by an exact diagonalization (ED) method. 
Since the system size is limited to being very small, 
however, 
the finite-size effect becomes significant. 
To avoid this difficulty as much as possible, 
Hida combined the critical point determined from the energy gap 
under the open boundary condition and the string correlation functions 
under the periodic boundary condition. 
For our purposes, we impose only the open boundary condition 
for our DMRG calculations and employ 
a definition of the string order Eq. (\ref{eq.5}) 
as the longest-ranged component. 

In the final stage of the present analysis, 
we estimate the critical exponents $\nu_\alpha(D_c, J^{\textrm{fix}}_z)$ 
and $\nu_\alpha(D^{\textrm{fix}}, J^{\textrm{c}}_z)$. 
We first consider the case where $D$ is controlled for a fixed $J_z$. 
Within the FSS analysis based on Eq. (\ref{eq.7}), 
an appropriate set of $D_{\textrm{c}}$, $\eta_\alpha$, and $\nu_\alpha$ is 
expected to give a universal function $\Psi$ near $D=D_c$ independent of $N$. 
However, it is difficult to determine the width of the critical region, 
as we have mentioned. 
We instead focus our attention on the gradient of the universal function $\Psi$ 
at $D=D_{\rm c}$. 
We note that the $N$-independence of the gradient is a necessary condition 
for the existence of the universal function $\Psi$ near $D=D_c$. 
Therefore, we assume that the gradient for $N=N_1$ and that for $N=N_2$ 
agree with each other for the same $D_{\textrm{c}}$, $\eta_\alpha$, and 
$\nu_\alpha$, to give 
\begin{eqnarray}
\left. N_1^{\eta_\alpha-1/\nu_\alpha} \frac{\partial [O^\alpha_{\textrm{str}}(N_1, D, J^{\textrm{fix}}_z)]}
       {\partial D} \right|_{D=Dc}   \hspace{10mm} \nonumber \\
= \left. N_2^{\eta_\alpha-1/\nu_\alpha} \frac{\partial [O^\alpha_{\textrm{str}}(N_2, D, J^{\textrm{fix}}_z)]}
       {\partial D} \right|_{D=Dc}
\label{eq.15-5}
\end{eqnarray}
We input $D_{\textrm{c}}$ and $\eta_\alpha(\tilde{N}, D, J_z) $ determined 
above into $D$ and $\eta$ in this equation and 
solve with respect to $1/\nu_\alpha$. 
Denoting the solution by 
$1/\nu_\alpha(\tilde{N}, D_c, J^{\textrm{fix}}_z)$, we obtain 
\begin{eqnarray}
\frac{1}{\nu_\alpha(\tilde{N}, D_c, J^{\textrm{fix}}_z)} = 
\eta_\alpha(\tilde{N}, D_{\textrm{c}}, J^{\textrm{fix}}_z) \hspace{20mm} \nonumber \\
+ \frac{\log[(O')^\alpha_{\textrm{str}}(N_2, D_{\textrm{c}}, J^{\textrm{fix}}_z)/
(O')^\alpha_{\textrm{str}}(N_1, D_{\textrm{c}}, J^{\textrm{fix}}_z) ]}{\log [N_2/N_1]},
\label{eq.16}
\end{eqnarray}
where $(O')^\alpha_{\textrm{str}}(N, D_{\textrm{c}}, J^{\textrm{fix}}_z)$ represents
\begin{align}
(O')^\alpha_{\textrm{str}}(N, D_{\textrm{c}}, J^{\textrm{fix}}_z) = 
\left. \frac{\partial O^\alpha_{\textrm{str}}(N, D, J_z^{\textrm{fix}})}
{\partial D} \right|_{D=D_{\textrm{c}}}.
\label{eq.17}
\end{align}
We note that $\nu_\alpha(\tilde{N}, D_c, J^{\textrm{fix}}_z)$ is a finite-size 
quantity and we examine the $N$-dependence of this quantity. 
An extrapolation to the limit $\tilde{N}\rightarrow\infty$ 
provides the exponent $\nu_\alpha(D_{\textrm{c}}, J^{\textrm{fix}}_z)$. 
Hereafter, we call $\eta_\alpha(\tilde{N}, D, J_z)$ and 
$\nu_\alpha(\tilde{N}, D_c, J^{\textrm{fix}}_z)$ 
the finite-size exponents. 

We next consider the case where $J_z$ is controlled for a fixed $D$. 
The same derivation as the above from Eq. (\ref{eq.9}) 
leads to the finite-size exponent: 
\begin{eqnarray}
\frac{1}{\nu_\alpha(\tilde{N}, D^{\textrm{fix}}, J^{\textrm{c}}_z)} = 
\eta_\alpha(\tilde{N}, D^{\textrm{fix}}, J^{\textrm{c}}_z) \hspace{20mm} \nonumber \\
+ \frac{\log[(O')^\alpha_{\textrm{str}}(N_2, D^{\textrm{fix}}, J^{\textrm{c}}_z)/
(O')^\alpha_{\textrm{str}}(N_1, D^{\textrm{fix}}, J^{\textrm{c}}_z) ]}{\log [N_2/N_1]},
\label{eq.18}
\end{eqnarray}
where $(O')^\alpha_{\textrm{str}}(N, D^{\textrm{fix}}, J^{\textrm{c}}_z)$ represents
\begin{align}
(O')^\alpha_{\textrm{str}}(N, D^{\textrm{fix}}, J^{\textrm{c}}_z) = 
\left. \frac{\partial O^\alpha_{\textrm{str}}(N, D^{\textrm{fix}}, J_z)}
{\partial J_z} \right|_{J_z=J_z^{\textrm{c}}}.
\label{eq.19}
\end{align}
The extrapolation of $\nu_\alpha(\tilde{N}, D^{\textrm{fix}}, J^{\textrm{c}}_z)$ 
gives the exponent $\nu_\alpha(D^{\textrm{fix}}, J^{\textrm{c}}_z)$.

\section{\label{sec:level4}Results and discussions}

\subsection{\label{sec:level4-1}Behavior of order parameters}
Let us first review the behavior of the four order parameters 
under consideration in a finite-size system and 
summarize some important relations between them. 
In a moderately large system, we can see indications of 
asymptotic behavior in each order parameter, although 
slow convergence prevents a full description.
Some are characteristic for a given 
region of the parameter space, which is specified as one of 
the Haldane, N$\acute{\textrm{e}}$el, Large-D, and $XY$ phases. 

\begin{figure}[htbp]
	\begin{center}
		\includegraphics[width=7.5cm]{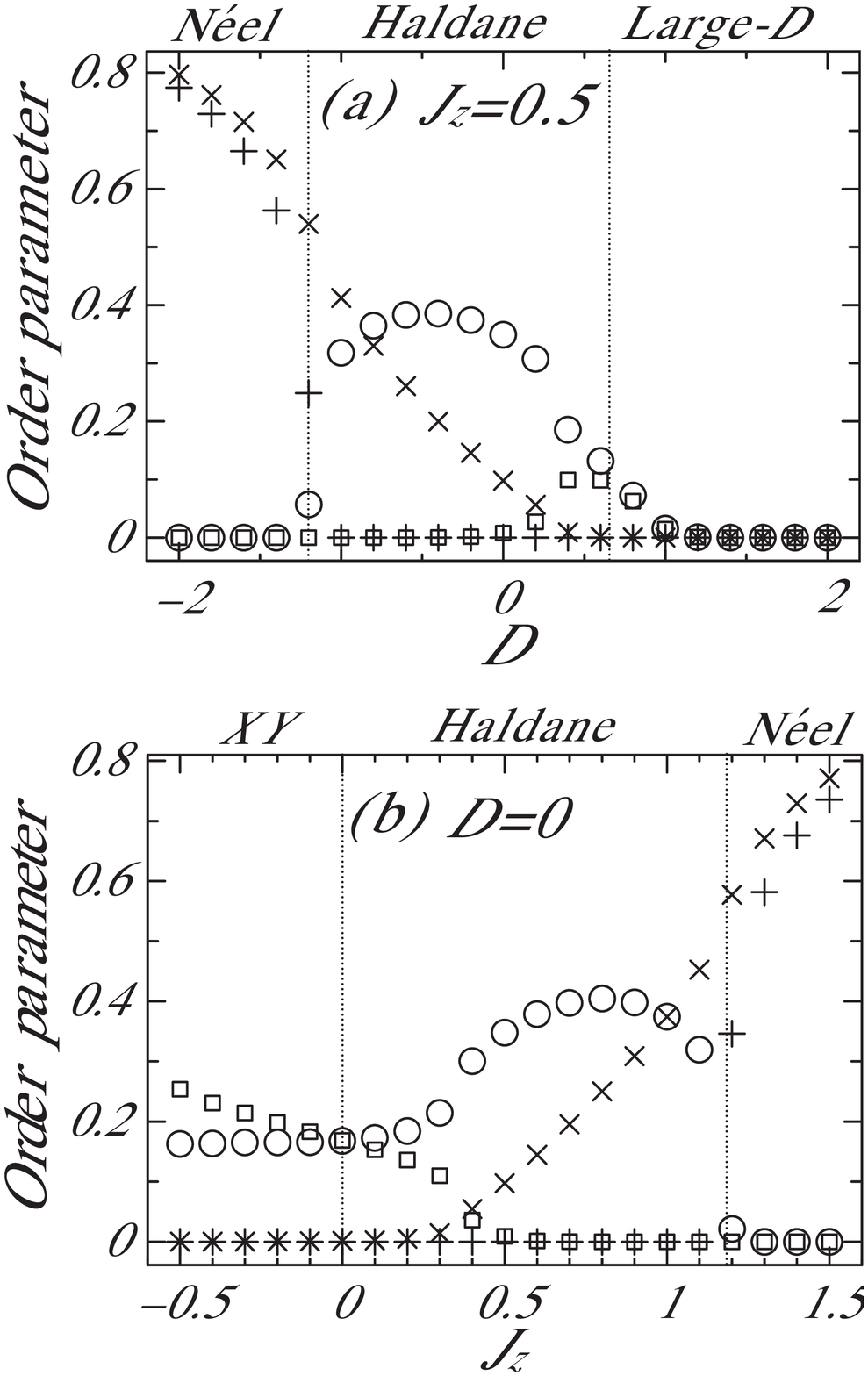}
	\end{center}
	\caption{
$(D, J_z)$ dependence of the order parameters 
for the Hamiltonian (\ref{eq.1}) on (a) the $J_z=0.5$ line 
and (b) the $D = 0$ line at $N = 300$. 
 $+$: $O^z_{\textrm{N}\acute{\textrm{e}}\textrm{el}}$, 
$\times$: $O^z_{\textrm{str}}$, 
$\square$: $O^x_{\textrm{N}\acute{\textrm{e}}{\textrm{el}}}$, 
$\bigcirc$: $O^x_{\textrm{str}}$.	}
	\label{fig.3}
\end{figure}

We illustrate $O^\alpha_{\rm str}(300, D, J_z)$ and 
$O^\alpha_{\rm N\acute{\textrm{e}}el}(300, D, J_z)$ with $\alpha=x$ or $z$ 
in Fig. \ref{fig.3}. 
The $D$-dependences of the order parameters on the line $J_z=0.5$ and 
their $J_z$-dependences on the line $D=0$ are shown 
in Fig. \ref{fig.3}(a) and \ref{fig.3}(b), respectively. 
We now compare the magnitudes of the four order parameters. 
(i) If $J_z>0$, $O^\alpha_{\textrm{str}}$ is larger 
than $O^\alpha_{\textrm{N}\acute{\textrm{e}}{\textrm{el}}}$. 
This is a known relation 
found by Kennedy and Tasaki \cite{Kennedy:CMP147}. 
(ii) When $D = 0$ and $J_z=1$, we have 
$O^x_{\textrm{str}}=O^z_{\textrm{str}}$ 
due to the isotropy of the system. 
(iii) When $J_z$ is decreased and 
when $J_z$ crosses a critical point at $J_z=0$, 
$O^x_{\textrm{str}}$ is smaller than 
$O^x_{\textrm{N}\acute{\textrm{e}}{\textrm{el}}}$. 
This fact will be discussed and utilized in \S \ref{sec:level4-4}. 
Although $O^x_{\textrm{N}\acute{\textrm{e}}\textrm{el}}$ appears to be nonzero 
around $0 < D < 1$ with $J_z = 0.5$ and 
around $0< J_z < 0.5$ with $D = 0$, 
we can confirm that 
$O^x_{\textrm{N}\acute{\textrm{e}}{\textrm{el}}}$ in this region 
vanishes for the long-ranged limit. 
On the other hand, $O^z_{\textrm{str}}$ around $0 < J_z < 0.4$ 
with $D = 0$ looks very small 
but it survives as a nonzero quantity 
in an infinite system, as shown in \S \ref{sec:level4}. 

The phase boundaries of Haldane--N$\acute{\textrm{e}}$el, 
Haldane--Large-$D$, and Haldane--$XY$ are denoted by dotted lines, though they are given only as indicators as we will determine 
the boundaries in the following subsections. 
We can see that 
some or all of the order parameters vanish at the phase boundaries. 
Also, the order parameters are continuous around the boundaries, 
which suggests that the phase transitions are continuous. 
Therefore, the FSS analysis and the GSPRG procedure are feasible 
for capturing critical phenomena 
in this case, except for the Berezinskii--Kosterlitz--Thouless (BKT) transition 
which does not satisfy the conditions of Eq. (\ref{eq.8a}) or (\ref{eq.10a}) 
and thus requires extra consideration. 
For GSPRG, however, it is possible for us to capture the transition 
by looking at the exponent $\eta_\alpha$ as discussed in 
\S \ref{sec:level4-4}. 

\begin{figure}[htbp]
	\begin{center}
	\includegraphics[width=7.5cm]{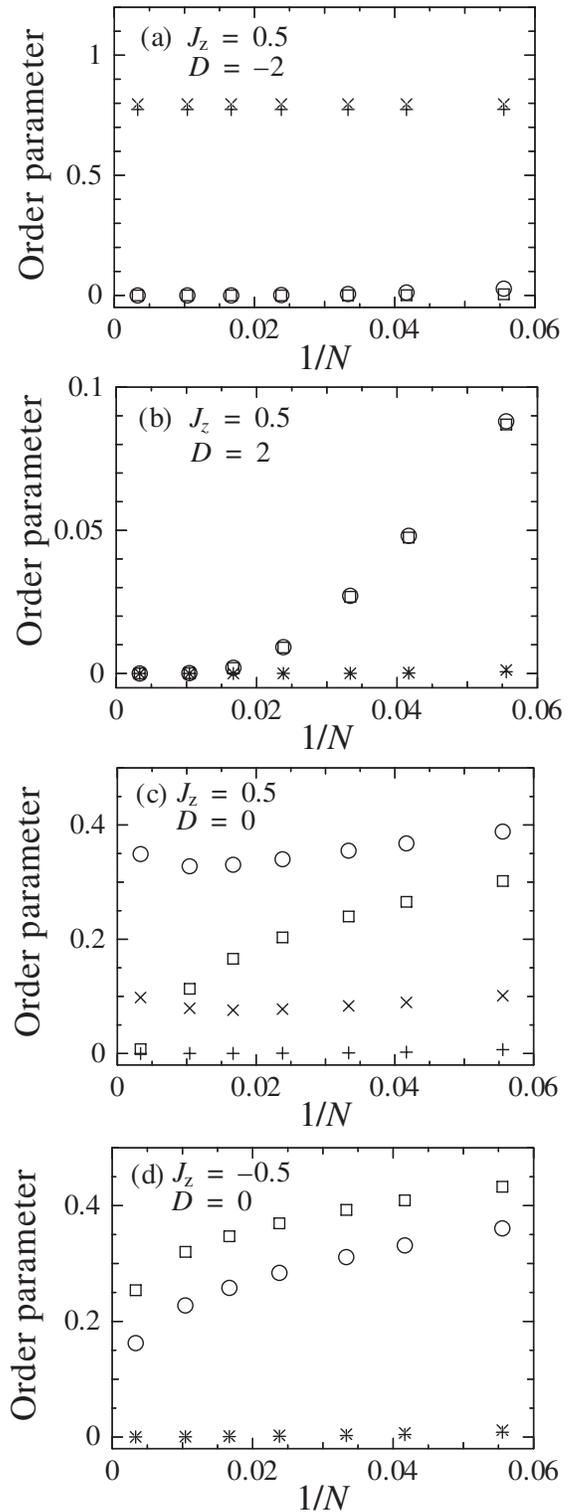}
	\end{center}
	\caption{
Order parameters for the Hamiltonian (\ref{eq.1}) as a function of 
the inverse of the system size. 
Each panel represents (a) the N$\acute{\textrm{e}}$el phase, (b) the Large-$D$ phase, 
(c) the Haldane phase, and (d) the $XY$ phase.
$+$: $O^z_{\textrm{N}\acute{\textrm{e}}{\textrm{el}}}$, 
$\times$: $O^z_{\textrm{str}}$, 
$\square$: $O^x_{\textrm{N}\acute{\textrm{e}}\textrm{el}}$, 
$\bigcirc$: $O^x_{\textrm{str}}$.
	}
	\label{fig.1}
\end{figure}

\begin{figure}[htbp]
	\begin{center}
	\includegraphics[width=7.5cm]{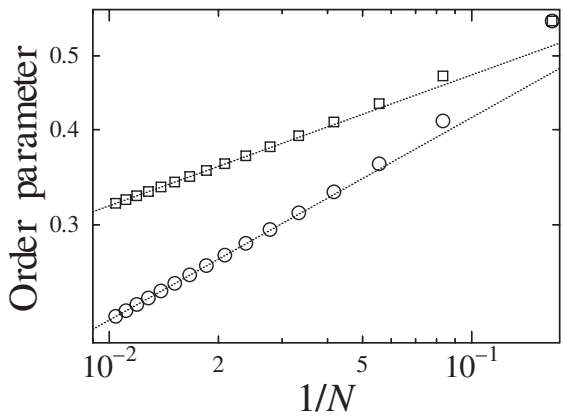}
	\end{center}
	\caption{
Order parameters for the Hamiltonian (\ref{eq.1}) 
in the $XY$ phase as a function of $1/N$. 
A fitting based on $O(N) = C_1N^{-\gamma}$  
is carried out for $N=96,90,\cdots,72$. 
$\square$: $O^x_{\textrm{N}\acute{\textrm{e}}\textrm{el}}$, 
$\bigcirc$: $O^x_{\textrm{str}}$.
	}
	\label{fig.2}
\end{figure}

We next observe the behavior of the four order parameters 
in each phase to determine their thermodynamic limits. 
In Fig. \ref{fig.1}, we illustrate 
the behavior of the order parameters (\ref{eq.5}) 
as a function of the inverse of the system size 
at the representative points 
$(D, J_z)=(-2, 0.5)$, $(2, 0.5)$, $(0, 0.5)$, and $(0, -0.5)$. 
These sets of parameters 
correspond to the N$\acute{\textrm{e}}$el phase, Large-$D$ phase, 
Haldane phase, and $XY$ phase, respectively. 
We observe that in the N$\acute{\textrm{e}}$el phase, 
only the order parameters in the $z$ direction remain nonzero 
in the limit $N\rightarrow\infty$. 
All the four parameters vanish in the Large-$D$ phase in the 
thermodynamic limit. 
We note that in the Haldane phase, 
only the string order parameters 
in the two directions remain nonzero in the thermodynamic limit. 
It is difficult to judge in Fig. \ref{fig.1}(d) whether 
both of the transverse order parameters in the $XY$ phase  
vanish or remain nonzero in the thermodynamic limit. 
We plot the same data on a logarithmic scale in Fig. {\ref{fig.2}}. 
For large $N$, the data exhibits a linear behavior,  
which suggests that 
the transverse order parameters in the $XY$ phase are critical, 
consistent with 
previous reports \cite{Alcaraz:PRB46, Hatsugai:PRB44}. 
Consequently we can confirm that 
the order parameter $O^\alpha(\infty, 0, -0.5)$ vanishes in the $XY$ phase. 
in the limit $N\rightarrow\infty$. 

\subsection{\label{sec:level4-2}Haldane--Large-$D$ transition line}
In this subsection, we examine the transition 
from the Haldane phase to the large-$D$ phase. 
This transition is known to be of Gaussian type. 
As we observe in \S \ref{sec:level4-1}, 
the string order parameters in the Haldane phase 
remain nonzero for both $\alpha=x$ and $z$ while 
both of the N$\acute{\textrm{e}}$el 
order parameters vanish along the directions 
$x$ and $z$. 
We also observe critical behavior near the transition line 
in both $O^x_{\textrm{str}}$ and $O^z_{\textrm{str}}$. 

\begin{figure}[htbp]
	\begin{center}
	\includegraphics[width=7.5cm]{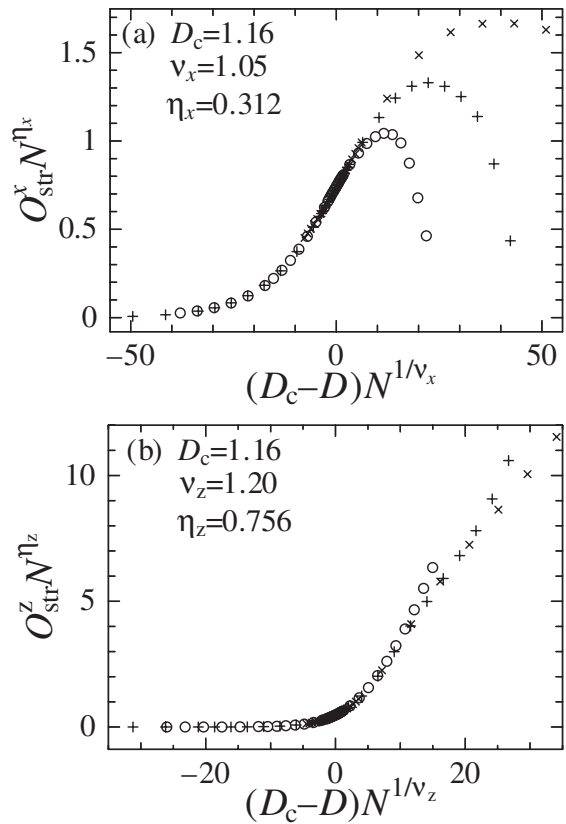}
	\end{center}
	\caption{
FSS of (a) the transverse string order and (b) the longitudinal string order on $J_z = 1.25$ near 
the Haldane--Large-$D$ critical point. 
$\bigcirc$, $+$, and $\times$ represent $N = 24, 48, 96$.
	}
	\label{fig.4}
\end{figure}

To begin with, we consider difficulties in 
the FSS analysis near the transition 
between the two phases. 
In this analysis, we have adjusted the critical point 
$D_{\rm c}$ and exponents $\eta_\alpha$ and $\nu_\alpha$ 
so that the data for $N=24$, 48, and 96 follows a universal function. 
The results are depicted in Fig. \ref{fig.4}. 
In Fig. \ref{fig.4}(a), we observe 
a deviation from the universal function $\Psi$ 
at $D$, not far from $D=D_{\rm c}$. 
The appearance of this deviation depends on 
the system size and the direction $\alpha$. 
Thus, it is not easy to determine the critical region 
around $D=D_{\rm c}$ with finite-size data 
less than 100 sites in this case.

Despite this difficulty, 
we can choose input parameters $D_{\rm c}$, $\nu_{\alpha}$, and 
$\eta_{\alpha}$ such that a universal function $\Psi$ appears 
near the transition point. 
In Fig. \ref{fig.4}(a), the string correlation functions 
in the direction $\alpha=x$ provide us with 
$D^{\textrm{HL}}_{\textrm{c}}(J_z=1.25) = 1.16$ and  
$\nu_x(D=1.16, J_z=1.25) = 1.05$. 
On the other hand, in Fig. \ref{fig.4}(b), 
the string correlation functions 
in $\alpha=z$ give 
$D^{\textrm{HL}}_{\textrm{c}}(J_z=1.25) = 1.16$ and  
$\nu_z(D=1.16, J_z=1.25) = 1.20$. 
The estimate of the transition point $D_{\rm c}$ for $\alpha=x$ 
and that for $\alpha=z$ agree with each other. 
This fact strongly suggests that 
the string correlation functions for the transverse and 
longitudinal directions reveal a common phase transition. 
We should note that 
$O^x_{\textrm{str}}$ and $O^z_{\textrm{str}}$ 
are clearly different quantities near the transition point, 
because there are differences in their exponents, for example 
$\eta_x=0.312$ and $\eta_z=0.756$. 
Our FSS analysis gives $\nu_x(D=1.16, J_z=1.25) = 1.05$ and 
$\nu_z(D=1.16, J_z=1.25) =1.20$. 
We recall that the growth of the correlation length 
determines the critical behavior near the transition point 
from a general argument of the renormalization group 
concerning critical phenomena. 
In this framework, only 
a single characteristic length in a system 
shows critical behavior. 
The characteristic length must be the correlation length 
of the system. 
Thus, the exponent of the correlation length 
should be unique for the order parameters. 
In this case, 
the correlation functions of the string order parameters 
along both $\alpha=x$ and $\alpha=z$ show critical behavior 
as shown by the FSS analysis. 
From this argument, $\nu_x$ and $\nu_z$ should exhibit 
a serious finite-size effect, which we will examine 
and solve by GSPRG analysis. 

\begin{figure}[htbp]
	\begin{center}
	\includegraphics[width=7.5cm]{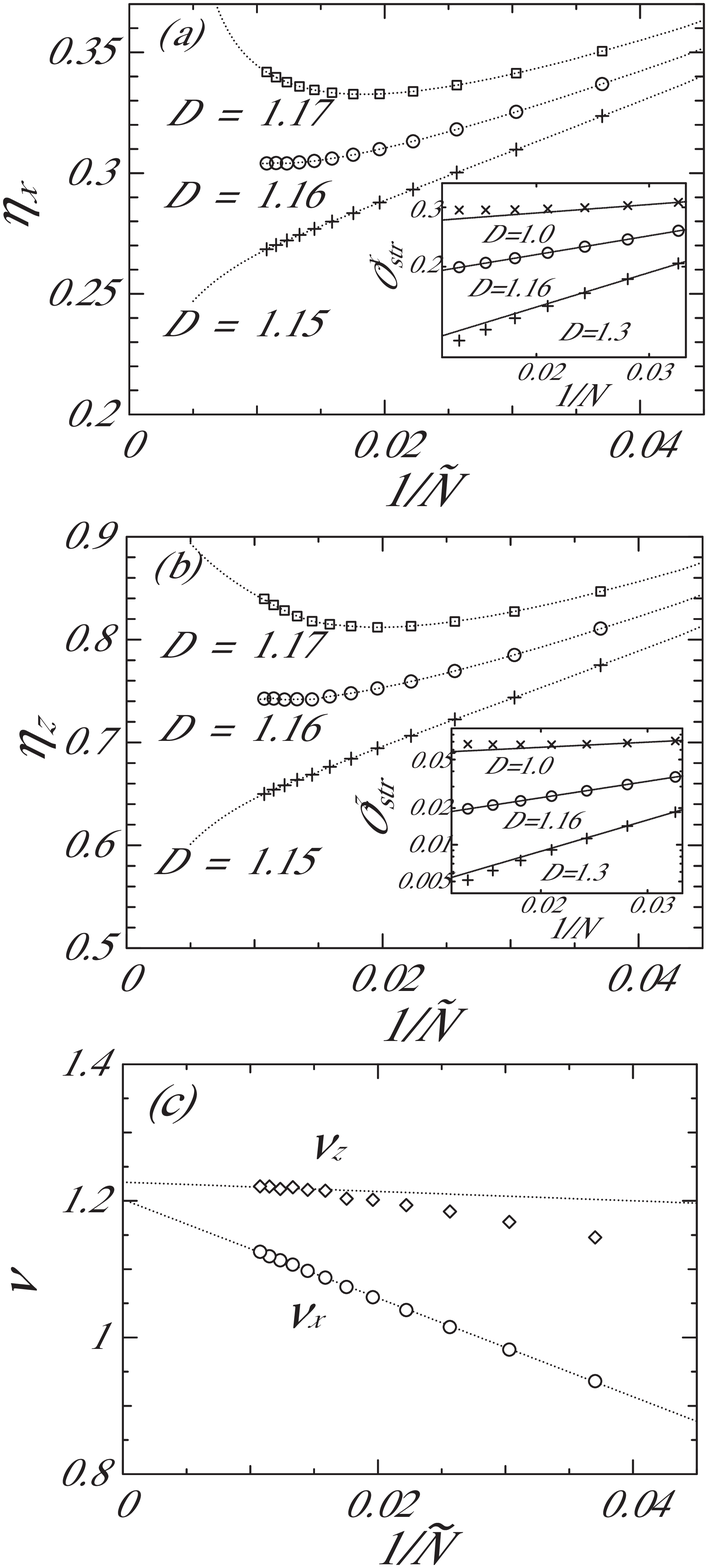}
	\end{center}
	\caption{
GSPRG analysis of (a) $\eta_x$, (b) $\eta_z$, 
and (c) $\nu_\alpha(D=1.16)$ 
on $J_z = 1.25$. $\tilde{N}$ is given by Eq. (\ref{eq.11}). 
The dotted curves in (a) and (b) are guides for the eyes. 
The dotted lines in (c) are linear fitting lines. 
The string correlation functions $O^x_{\rm str}(N, D, J_z=1.25)$ and 
$O^z_{\rm str}(N, D, J_z=1.25)$ as a function of $1/N$ are shown in 
the logarithmic scale in the inset figures of 
(a) and (b), respectively. 
	}
	\label{fig.5}
\end{figure}

We consider the case of $J_z=1.25$ in order to observe the finite-size effect. 
In Fig. \ref{fig.5} we illustrate our results for the exponents 
$\eta_x(\tilde{N},D,1.25)$, $\eta_z(\tilde{N},D,1.25)$, 
$\nu_x(\tilde{N},1.16,1.25)$, and $\nu_z(\tilde{N},1.16,1.25)$ 
determined by GSPRG analysis. 
In Fig.\ref{fig.5}(a), 
we observe the critical-disordered boundary 
at $D=1.17, N'=55.3$. 
The critical-ordered boundary is also observed 
at $D=1.15, N'=45.5$. 
On the other hand, we obtain no boundaries defined
in eqs. (\ref{co-boundary1})-(\ref{cd-boundary2}) 
in the case of $D=1.16$. This fact suggests that 
the critical region for finite-size systems in our study 
is realized around $D=1.16$ with a narrow width. 
In order to confirm whether the width shrinks or not 
as the system sizes increase, 
we examine the relationship between $N^{\prime}$ and $D$ 
so that the case is on the boundary. 
We obtain some of the critical-disordered boundaries 
at $(D=1.19, N'=36.8)$, $(D=1.2, N'=32.7)$, 
and $(D=1.21, N'=29.8)$. 
We also obtain some of the critical-ordered boundaries 
at $(D=1.14, N'=32.7)$, $(D=1.13, N'=30.0)$, 
and $(D=1.12, N'=29.1)$. 
These results indicate that the critical region for a given 
$N^{\prime}$ is gradually narrower when $N^{\prime}$ increases 
although the expression of the relationship 
between $N^{\prime}$ and $D$ on the boundary is unknown 
in the present stage. 
It is reasonable to conclude that the critical region shrinks and 
goes to the transition point for the infinite-size system.
When one can confirm whether the critical region 
between the two boundaries is sufficiently narrow or not, 
the width of the region should be regarded as an error 
coming from the maximum system size and the interval of $D$ 
in the performed calculations.
In this work, thus, we conclude 
$D^{\textrm{HL}}_{\textrm{c}}(J_z=1.25) = 1.16 \pm 0.01$. 
Note here that we can obtain the same critical point 
from $\eta_{z}$ in Fig.\ref{fig.5}(b) in the same manner. 
Hereafter, we determine critical points with an error 
in this way. 
In order to confirm whether the critical behavior (\ref{eq.8b}) 
or (\ref{eq.10b}) appears or not in the original correlation 
functions, each string correlation function as a function of 
$1/N$ is shown in the logarithmic scale in inset figures. 
The finite-size string correlations for each direction 
clearly reveal a power-law decay behavior
at the critical point $D^{\textrm{HL}}_{\textrm{c}}(J_z=1.25) = 1.16$. 
On the other hand, a behavior deviating from power-law decay 
appears in the cases of $D=1.0$ and $D=1.3$ in the ordered and disordered phases, respectively, 
as we have mentioned in \S \ref{sec:level3-3}. 
Note here that a comparison with these insets 
shows that the system size dependence of 
the finite-size quantity (\ref{eq.11}) sensitively change 
near the transition point. 
We next observe the $\tilde{N}$ dependence 
of the finite-size exponents of $\nu_x(\tilde{N},1.16,1.25)$ and 
$\nu_z(\tilde{N},1.16,1.25)$ 
for $J_z=1.25$ and $D^{\textrm{HL}}_{\textrm{c}}(J_z=1.25)=1.16$ 
in Fig.\ref{fig.5}(c). 
These two finite-size exponents, $\nu_x$ and $\nu_z$, 
get gradually closer with increasing $\tilde{N}$. 
In the limit $\tilde{N}\rightarrow\infty$, 
$\nu_x$ and $\nu_z$ appear to approach a single value $\sim$1.2. 
This is consistent with the above argument 
on the unique characteristic length. 
Consequently, the problem of the disagreement of 
$\nu_x$ and $\nu_z$ in the FSS analysis occurs due to the 
finite-size effect and is resolved by GSPRG analysis.  

We now consider the transition point $D_{\rm c}$ 
for a fixed $J_z=1$. 
In this case, many studies have reported various estimates 
for the boundary of the Haldane phase, $D_{\rm c}$: 
$D_{\rm c}=0.93\pm 0.02$ in Ref. \cite{Sakai_Takahashi:PRB42}, 
$D_{\rm c}=0.99\pm 0.02$ in Ref. \cite{Golinelli:PRB46}, 
$D_{\rm c}=0.90\pm 0.05$ in Ref. \cite{Tonegawa:JPSJ65}, 
$D_{\rm c}=1.001\pm 0.001$ in Ref. \cite{Chen:JPSJ69}, 
$D_{\rm c}=0.95$ in Ref. \cite{Koga:PL296}, 
$D_{\rm c}=0.99$ in Ref. \cite{Boschi:EPJB35}, and 
$D_{\rm c}\sim0.97$ in Ref. \cite{Tzeng:PRA77}. 
Among these works, only a single study \cite{Tonegawa:JPSJ65} 
was based on the analysis of the string order, although 
data from the numerical-diagonalization calculations in this study 
for small clusters might not be sufficient to show the transition point. 
Recently, Tzeng and Yang \cite{Tzeng:PRA77} 
investigated the fidelity susceptibility \cite{Venuti:PRL99} 
of the ground state by the DMRG method 
to detect quantum phase transitions for the system. 
This work 
examines only the information of the ground state,  
a feature that is shared with our present analysis. 
Other works analyzed the structure of low-energy levels. 
From the present analysis, our estimate is 
$D_{\rm c}=0.975\pm 0.015$, which we have obtained 
irrespective of $\alpha=x$ or $\alpha=z$. 
Although the estimates are all very close to each other, 
there are small differences between them 
even taking errors into account.  
The reason for these differences is not clear at present 
and should be resolved as a future issue. 

Next, we consider the transition point $D_{\rm c}$ 
for a fixed $J_z=0.5$. 
The estimation of this point is suitable for checking 
the availability of our analysis procedure, 
because a relatively large exponent $\nu$ 
which is reported 2.38 by analysis of the energy level 
structure appears \cite{Boschi:EPJB35}.
Several previous studies presented numerical data 
of the transition point as follows: 
$D_{\rm c}=0.635$ in Ref. \cite{Hida:PRB67}, 
$D_{\rm c}=0.65$ in Ref. \cite{Boschi:EPJB35}, 
$D_{\rm c}=0.633 \pm 0.02$ in Ref. \cite{Roncaglia:PRB77}. 
All of these works examined the free energy 
near the critical point to determine the critical point. 
In particular, the recent study \cite{Roncaglia:PRB77} 
develops rapidly converging methods by using 
the differentiations of a quantity, which is derivative of 
the ground state energy with respect to a controlled parameter, 
as a function of $N$. 
From the viewpoint of using only information 
in the ground state for detecting a quantum phase transition, 
our analysis and their analysis have a common policy.
Our estimate for $D_{\rm c}(J_z = 0.5)$ is $0.67 \pm 0.04$, and this estimate 
is also consistent with all previous reports. 

\begin{figure}[htbp]
	\begin{center}
	\includegraphics[width=7.5cm]{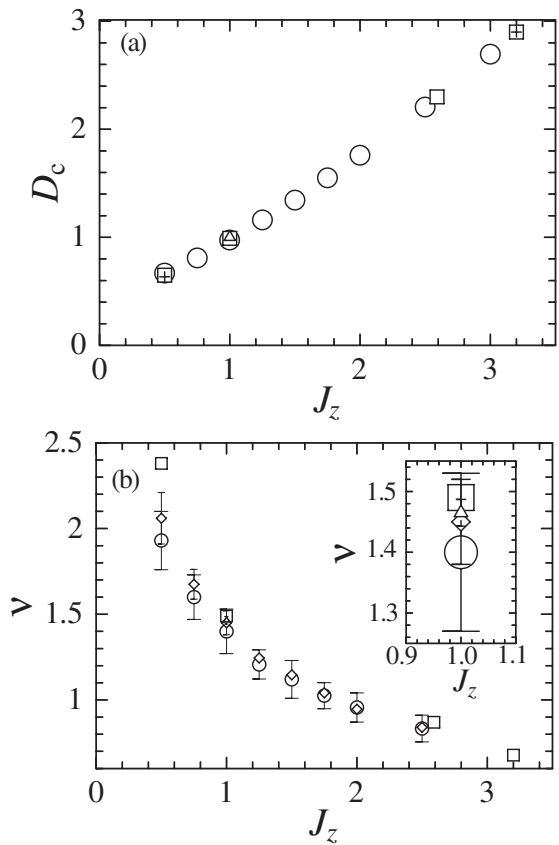}
	\end{center}
	\caption{
(a) Haldane--Large-$D$ transition line and 
(b) critical exponent $\nu_\alpha(J_z)$. 
$\bigcirc$: evaluated value of x-component, 
$\Diamond$: evaluated value of z-component, 
+: Ref. \cite{Hida:PRB67}, 
$\square$: Ref. \cite{Boschi:EPJB41}, 
$\bigtriangleup$: Ref. \cite{Chen:JPSJ69}. 
The inset of (b) magnifies the data at $J_z = 1$ to allow a clear comparison to distinguish the data. 
	}
	\label{fig.6}
\end{figure}

In accordance with the above results, 
we apply the procedure to estimate the critical behavior 
for other $J_z$, confirming the $J_z$ dependence of 
$D^{\textrm{HL}}_{\textrm{c}}(J_z)$ and $\nu(D^{\textrm{HL}}_{\textrm{c}}, J_z)$. 
The error of $\nu(D^{\textrm{HL}}_{\textrm{c}}, J_z)$ is estimated by 
$|\nu(D^{\textrm{HL}}_{\textrm{c}}, J_z) - \nu(D^{\tilde{N} = 93, \textrm{HL}}_{\textrm{c}}, J_z)|$. 
We illustrate our results in Fig. \ref{fig.6} together with 
those of previous reports \cite{Chen:JPSJ69, Hida:PRB67, 
Boschi:EPJB41}. 
Our estimates of $D^{\textrm{HL}}_{\textrm{c}}(J_z)$ and 
$\nu(D^{\textrm{HL}}_{\textrm{c}}, J_z)$ are common 
for $\alpha=x$ and $\alpha=z$ within errors. 
Our transition line is almost consistent 
with those of previous reports \cite{Chen:JPSJ69, Hida:PRB67, 
Boschi:EPJB41}, 
in which the energy-level structure is analyzed. 
Our estimates of $\nu$ also agree well 
with previous reports within errors. 
Consequently, our GSPRG analysis successfully captures the transition 
between the Haldane phase and the large-$D$ phase. 

\begin{figure}[htbp]
	\begin{center}
	\includegraphics[width=7.5cm]{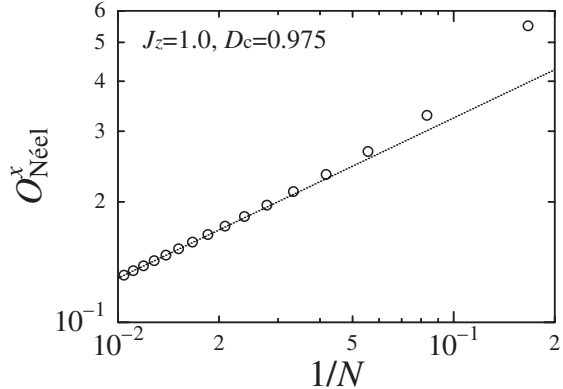}
	\end{center}
	\caption{$1/N$ dependence of 
$O_{\rm N\acute{\textrm{e}}el}^{x}(N,D,J_{z})$ 
at $J_{z}=1$ and $D_{\rm c}=0.975$. 
The dotted line shows 
$O_{\rm N\acute{\textrm{e}}el}^{x}(N,D,J_{z})\propto 
N^{-\eta^{x}_{\rm N\acute{\textrm{e}}el}}$ 
with $\eta^{x}_{\rm N\acute{\textrm{e}}el}=0.40$. 
	}
	\label{sxsx_cor}
\end{figure}

The correlation length exponent $\nu$ is known to 
be related to other critical exponents. 
In the Gaussian transition, Okamoto obtained 
the following relationship 
from the argument by the bosonization method:  
\begin{equation}
\nu= \frac{2}{4-\eta^{z}_{\rm N\acute{\textrm{e}}el}},
\label{Okamoto_relation_nu_eta}
\end{equation}
where $\eta^{\alpha}_{\rm N\acute{\textrm{e}}el}$ is 
the exponent defined by 
$\langle S_{0}^{\alpha}S_{r}^{\alpha}\rangle\sim (-1)^{r} 
r^{-\eta^{\alpha}_{\rm N\acute{\textrm{e}}el}}$ 
at the transition point. 
Note here that 
$\eta^{x}_{\rm N\acute{\textrm{e}}el}
\eta^{z}_{\rm N\acute{\textrm{e}}el}=1$ holds. 
To confirm the consistency between 
our estimate of $\nu$ and the decay of the 
N$\acute{\textrm{e}}$el correlation function, 
we plot our $O_{\rm N\acute{\textrm{e}}el}^{x}(N,D,J_{z})$ 
at $J_{z}=1$ and $D_{\rm c}=0.975$ 
as a function of $1/N$ on a logarithmic scale 
in Fig. \ref{sxsx_cor}.  
We clearly observe a linear behavior for large $N$. 
We have added the dotted line 
$O_{\rm N\acute{\textrm{e}}el}^{x}(N,D,J_{z})\propto 
N^{-\eta^{x}_{\rm N\acute{\textrm{e}}el}}$ 
with $\eta^{x}_{\rm N\acute{\textrm{e}}el}=0.40$. 
From Eq. (\ref{Okamoto_relation_nu_eta}), 
this value of $\eta^{x}_{\rm N\acute{\textrm{e}}el} 
(=1/\eta^{z}_{\rm N\acute{\textrm{e}}el})$ 
gives $\nu \sim 1.33$, which is consistent with 
our estimate shown in the inset of Fig. \ref{fig.6}. 
This consistency also supports the scaling hypothesis 
that the growth of the unique correlation length 
determines all the critical behavior 
around the transition point. 

\subsection{\label{sec:level4-3}Haldane--N$\acute{\textrm{e}}$el transition line}
In this subsection, we examine 
the transition from the Haldane phase 
to the N$\acute{\textrm{e}}$el phase. 
This transition is considered to be of Ising type. 
We recall that 
in a transition of Ising type, 
the exponent of the correlation length 
is $\nu=1$ when the system approaches the transition point. 

We have mentioned in the above that 
the longitudinal string order is nonzero 
in both of the Haldane phase and 
the N$\acute{\textrm{e}}$el phase 
and that the order does not reveal the critical behavior 
at the transition point. 
This means that the longitudinal string order is not 
appropriate for studying the Haldane--N$\acute{\textrm{e}}$el 
transition. 
Therefore, to study this transition we examine only the transverse string order. 
By GSPRG analysis of this order,  
we determine the transition point $J_z^{\textrm{c}}$ 
for a given $D$ or 
the transition point $D_{\rm c}$ for a given $J_z$ 
and the critical exponent $\nu$ near the transition point. 

\begin{figure}[htbp]
	\begin{center}
	\includegraphics[width=7.5cm]{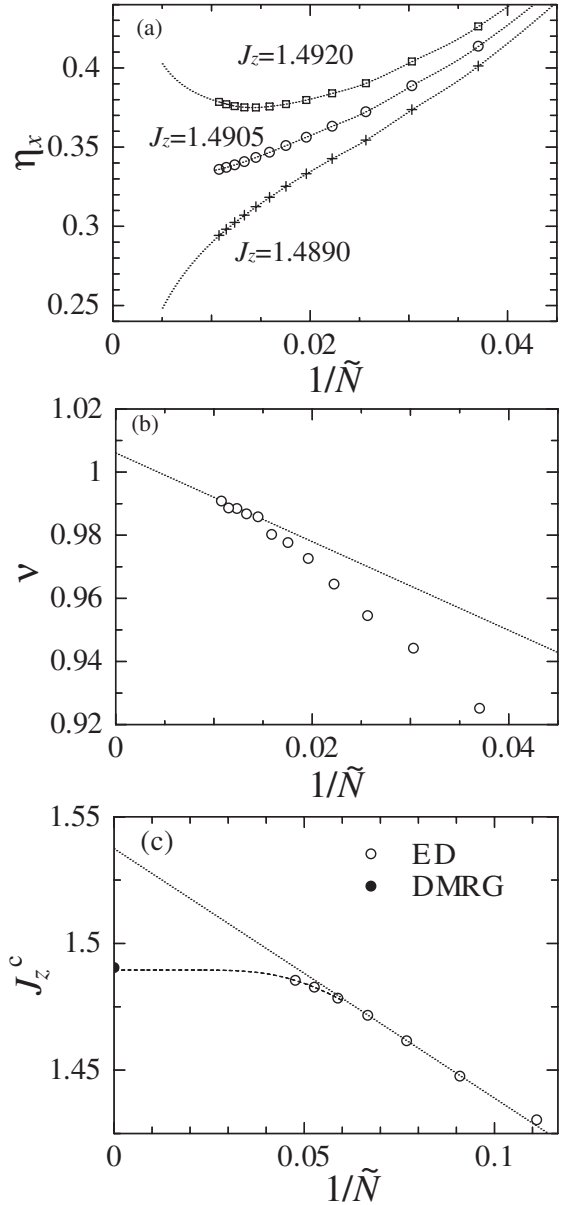}
	\end{center}
	\caption{
GSPRG analysis of (a) $\eta_x$ and (b) $\nu(J_z=1.4905)$ 
for $D = 0.5$ from our DMRG results for $\Delta N=6$. 
The dotted curves in (a) are guides for the eyes. 
The dotted line in (b) is a linear fitting line 
applied to the data for large $\tilde{N}$.  
(c) Extrapolation procedure of the finite-size critical point 
$J^{\textrm{c}}_z$ for $D = 0.5$ 
by the numerical-diagonalization method 
for $\Delta N=2$. 
The numerical-diagonalization data for $\tilde{N} =17, 19, 21$ 
are new in this paper. 
The dotted line is the extrapolation line 
in Ref. \cite{Hida:PRB67}. 
The broken curve is a guide for the eyes. 
	}
	\label{fig.7}
\end{figure}

We consider the case of $D=0.5$. 
We illustrate our result for finite-size exponents $\eta_x$ 
and $\nu$ in Fig. \ref{fig.7}(a) and (b), respectively.
Our estimates are 
$J^{\textrm{c,HN}}_{z}(D=0.5)=1.4905 \pm 0.0015$ 
and $\nu(J_{z}=1.4905, D=0.5) = 1.006 \pm 0.016$.  
Our estimate of the transition point is different 
from that of $J^{\textrm{c,HN}}_z(D=0.5)=1.536$ reported 
in Ref. \cite{Hida:PRB67}. 
To find the reason for the difference 
between the two estimates, we have made  
numerical-diagonalization calculations 
of finite-size clusters up to $N = 22$ 
under the periodic boundary condition 
and obtained the eigenenergies of the low-energy states. 
We have performed the same analysis as that 
in Ref. \cite{Hida:PRB67} and determined 
the finite-size critical point $J^{\textrm{c,HN}}_z(\tilde{N}, D=0.5)$ 
as $J_z$ at which the scaled energy gap does not depend 
on the system size for $\Delta N=2$. 
The results are depicted in Fig. \ref{fig.7}(c). 
From our numerical data for $N=8, 10, \cdots, 16$, 
we successfully reproduce the results 
of Ref. \cite{Hida:PRB67}. 
On the other hand, we can observe that 
$J^{\textrm{c,HN}}_z(\tilde{N}, D=0.5)$ of 
$\tilde{N} = 17, 19, 21$ gradually departs from 
the fitting line of the extrapolation 
in Ref. \cite{Hida:PRB67}. 
Our new data points approach our estimates from 
the string order by the DMRG calculations, 
as shown by the guide for the eyes denoted 
by the broken curve in Fig. \ref{fig.7}(c). 
This agreement suggests that the results from the 
numerical-diagonalization and DMRG calculations are 
consistent with each other 
if we accept the interpretation suggested by the broken curve.
Hence, careful extrapolation with respect to system size 
is required. 

The $\tilde{N}$ dependence of our new data appears exponential 
rather than polynomial. 
A similar $\tilde{N}$ dependence 
of $J^{\textrm{c,HN}}_z(\tilde{N}, D=1)$ 
was reported in Ref. \cite{Boschi:EPJB41}, 
in which calculations up to $N = 48$ based 
on the multi-target DMRG method 
with an infinite-system algorithm were carried out 
under the periodic boundary condition. 
Our result and Ref. \cite{Boschi:EPJB41} suggest that 
the absence of polynomial components 
does not depend on the values of the parameters of the system. 
It is important to be careful when a system-size extrapolation 
of an Ising transition point is carried out 
by the PRG analysis of the energy-level structure. 

We now compare estimates of the transition point 
between Ref. \cite{Boschi:EPJB41} and the present analysis. 
Reference \cite{Boschi:EPJB41} gives 
$J^{\textrm{c, HN}}_z(D=0)=1.186$. 
From the present analysis of our data up to $N = 150$, 
we obtain $J^{\textrm{c, HN}}_z(D=0) = 1.1860 \pm 0.0003$ 
for the transition point. 
Our estimate, with a very small error, agrees excellently 
with the estimate in Ref. \cite{Boschi:EPJB41}. 

We now discuss our estimate of $\nu$. 
Our estimate $\nu(J_{z}=1.4905, D=0.5) = 1.006 \pm 0.016$ 
is in good agreement with $\nu=1$ of the Ising-type transition. 
This agreement also suggests that 
our analysis successfully captures 
the Haldane--N$\acute{\textrm{e}}$el transition 
as well as the Haldane--Large-$D$ transition. 

\begin{figure}[htbp]
	\begin{center}
	\includegraphics[width=7.5cm]{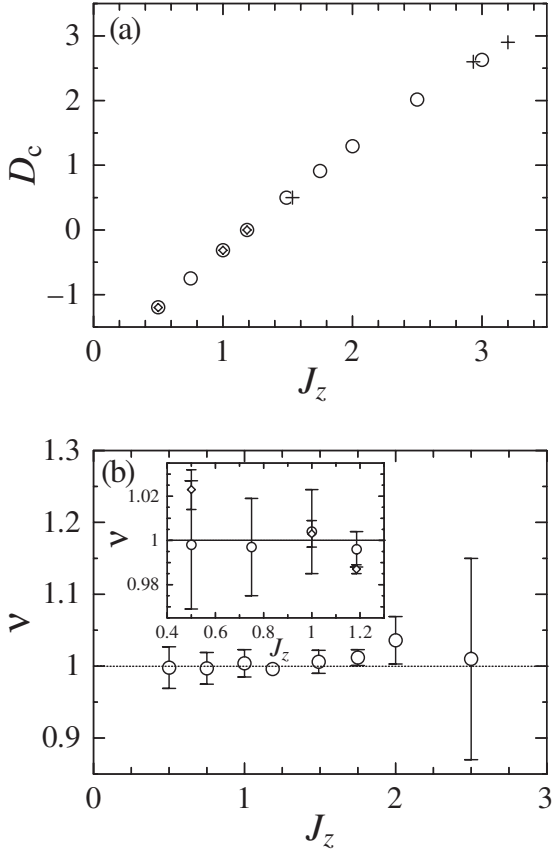}
	\end{center}
	\caption{
(a) Haldane--N$\acute{\textrm{e}}$el transition points. $\bigcirc$: Our work, 
+: Ref. \cite{Hida:PRB67}, $\Diamond$: Ref. \cite{Boschi:EPJB41}.
(b) Haldane--N$\acute{\textrm{e}}$el critical exponent $\nu$.
The inset figure compares our $\nu$ with that of Ref. \cite{Boschi:EPJB41}.
	}
	\label{fig.8}
\end{figure}

We can now summarize 
our results for the transition points $D_{\rm c}$ 
for a given $J_z$ 
and the critical exponents $\nu$ between the Haldane and 
the N$\acute{\textrm{e}}$el phases from our DMRG data. 
The results are depicted in Fig. \ref{fig.8}. 
Figure \ref{fig.8}(a) shows  
that our estimates for the transition points 
are in good agreement 
with the results in Ref. \cite{Boschi:EPJB41} 
of the multi-target DMRG method and 
the results in Ref. \cite{Hida:PRB67} 
of the numerical diagonalizations. 
In Fig. \ref{fig.8}(b),  
our estimates for the exponent agree well with $\nu=1$ 
irrespective of $J_z$. 
Note here that the center values of our estimates, 
namely the extrapolated results, are much closer to $\nu=1$ 
than the results in Ref. \cite{Boschi:EPJB41}, 
although our errors are estimated to be larger. 
Note also that 
the error in $\nu(D^{\textrm{HN}}_{\textrm{c}}=2.015, J_z=2.5)$ 
is quite large. 
The reason for this is considered to be that the curve 
of the Haldane--N$\acute{\textrm{e}}$el transition points and 
that of the Haldane--Large-$D$ transition points approach 
each other. 
A similar phenomenon appears when 
the central charge $c$ on the curve of 
the Haldane--Large-$D$ transition points 
was estimated in Ref. \cite{Hida:PRB67}, 
in which the estimate of $c$ gradually deviates from $c=1$ 
around $J_z \gtrsim 1$. 
In the report of Tzeng and Yang \cite{Tzeng:PRA77}, 
the transition point and the critical exponent are given as 
$Dc^{\rm HN}(J_z = 1) \sim -0.31$, 
$\nu(D^{\textrm{HN}}_{\textrm{c}}=-0.31, J_z=1.0) \sim 1.05$, respectively, 
from fidelity susceptibility analysis. 
Our estimated values at the same point are $Dc = -0.315 \pm 0.003$, 
$\nu = 1.004 \pm 0.019$, which are more precise than the values of Tzeng and Yang. 

\subsection{\label{sec:level4-4}Haldane--$XY$ transition line}
In this subsection, we examine the transition 
from the Haldane phase to the $XY$ phase. 
This transition is considered to be a BKT-type transition. 
We recall that in a BKT-type transition, 
the exponents $\eta_{x}=1/4$ and $\eta_{z}=1$ 
appear at the transition point 
and the exponent $\nu$ cannot be defined 
because the correlation length grows exponentially. 

\begin{figure}[htbp]
	\begin{center}
	\includegraphics[width=7.5cm]{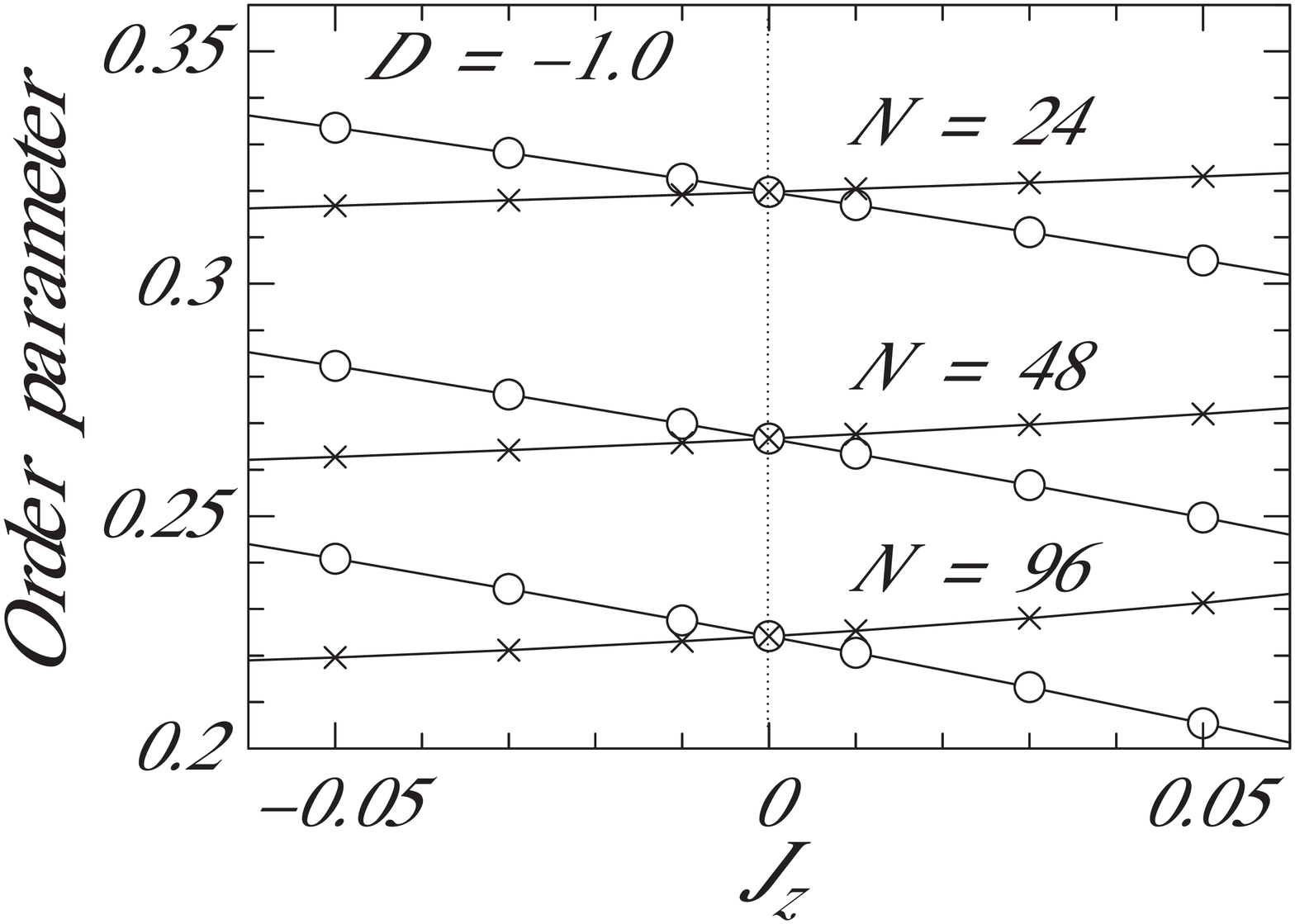}
	\end{center}
	\caption{
Order parameters crossing at $D=-1.0$. $\bigcirc$ and $\times$ represent 
$O^x_{\textrm{N}\acute{\textrm{e}}\textrm{el}}$ and $O^x_{\textrm{str}}$.
	}
	\label{fig.9}
\end{figure}

We consider the case $J_z<0$ and 
examine the magnitudes of the string order and 
the N$\acute{\textrm{e}}$el order. 
We refer back to the behavior of orders characterizing 
the Haldane phase, in which 
we have 
\begin{equation}
|O^{z}_{\textrm{str}}| > 0
\mbox{ and }
|O^{x}_{\textrm{str}}| > 0, 
\label{inequality_str}
\end{equation}
under the condition 
\begin{equation}
O^{x}_{\textrm{N}\acute{\textrm{e}}\textrm{el}} 
=O^{z}_{\textrm{N}\acute{\textrm{e}}\textrm{el}}= 0.
\label{vanishing_neel}
\end{equation}
This means that 
the region 
\begin{equation}
|O^\alpha_{\textrm{str}}| \le 
|O^{\alpha}_{\textrm{N}\acute{\textrm{e}}\textrm{el}}|,
\label{inequality_orders}
\end{equation} 
is not in the Haldane phase 
because 
the inequality (\ref{inequality_str}) and 
Eq. (\ref{vanishing_neel}) cannot both be satisfied at the same time 
assuming the inequality (\ref{inequality_orders}). 
However, it is not as easy to make 
a direct comparison of these quantities 
in the limit $N\rightarrow\infty$ as for 
the inequality (\ref{inequality_orders}). 
We can instead compare 
the finite-size quantities 
$|O^\alpha_{\textrm{str}}(N, D, J_z)|$ and 
$|O^\alpha_{\textrm{N}\acute{\textrm{e}}\textrm{el}}(N, D, J_z)|$. 
Recall that for $N=300$, 
$|O^\alpha_{\textrm{str}}(N, D, J_z)|$ is smaller than 
$|O^\alpha_{\textrm{N}\acute{\textrm{e}}\textrm{el}}(N, D, J_z)|$ 
when $J_z<0$, 
whereas 
$|O^\alpha_{\textrm{str}}(N, D, J_z)|$ is larger than 
$|O^\alpha_{\textrm{N}\acute{\textrm{e}}\textrm{el}}(N, D, J_z)|$ 
when $J_z >0$. 
We have studied the system size dependence of this behavior;
our results are depicted in Fig. \ref{fig.9}. 
The behavior is clearly independent of system size. 
We can also confirm this independence irrespective of $D$ 
for cases between the Haldane phase and the $XY$ phase. 
Our present results suggest 
the inequality (\ref{inequality_orders}) 
and indicate that the Haldane--$XY$ transition point satisfies 
$J^{\textrm{c,HXY}}_z(D) \ge 0$.  
The finding is entirely consistent 
with previous works. 
Thus, it is sufficient to consider the case of $J_z \ge 0$ 
hereafter in examining the Haldane--$XY$ transition. 

\begin{figure}[htbp]
 	\begin{center}
 	\includegraphics[width=8cm]{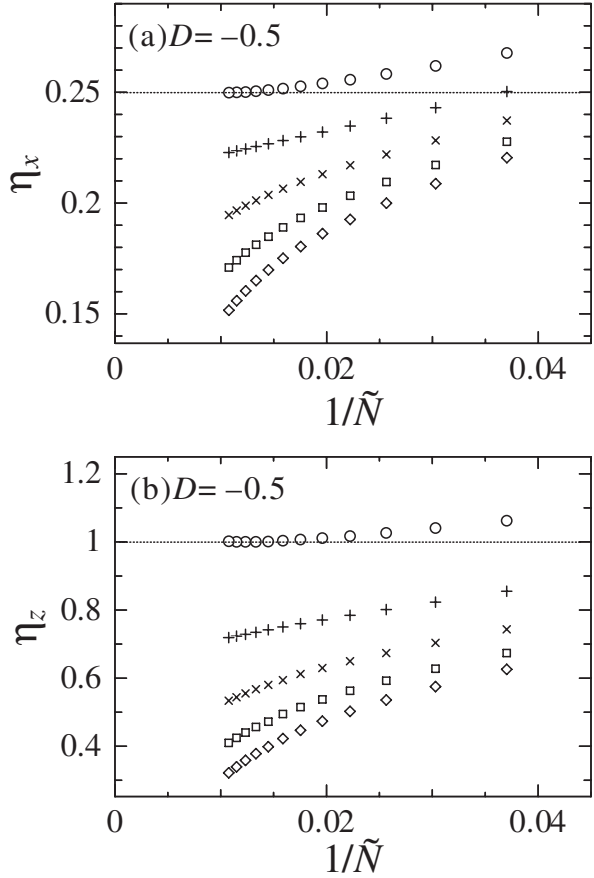}
 	\end{center}
 	\caption{
 GSPRG analysis of the exponent $\eta_\alpha(\tilde{N}, D = -0.5, J_z = 0)$. 
 The dotted lines are the rigorous exponent values of the BKT transition: 
 $\eta_x = 0.25$, $\eta_z = 1$. 
 $\bigcirc$, +, $\times$, $\square$, and $\Diamond$ represent 
 $J_z = 0$, 0.1, 0.15, 0.18, and 0.2, respectively.
 	}
 	\label{fig.10}
 \end{figure}

\begin{figure}[htbp]
	\begin{center}
	\includegraphics[width=7.5cm]{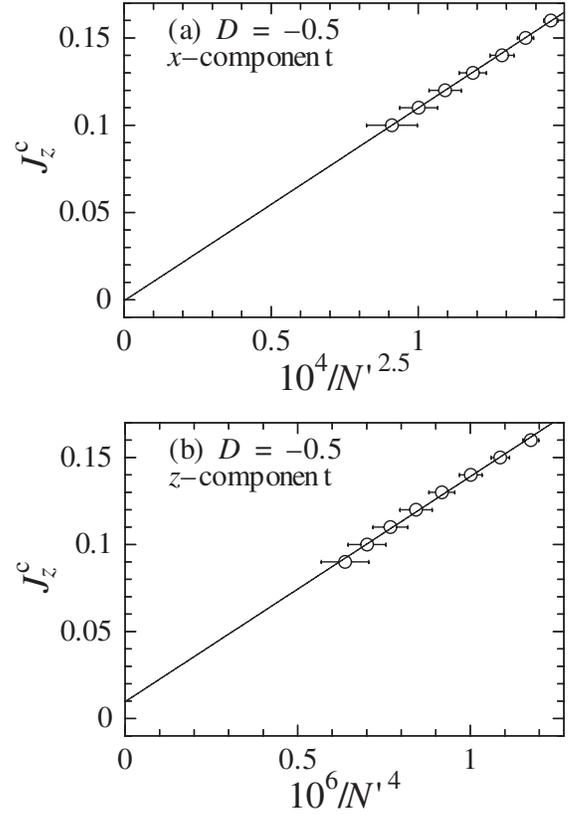}
	\end{center}
	\caption{
Behavior of the critical-ordered boundary point $J^{\textrm{c, HXY}}_z$ at $D=-0.5$ 
as a function of the system size $N'$. 
Results are shown for $m_{\textrm{F}} = 150$. 
The error is estimated from the difference between 
the results for $m_{\textrm{F}} = 100$ and 
for $m_{\textrm{F}} = 150$. 
	}
	\label{fig.11-5}
\end{figure}

 We now estimate $J^{\textrm{c,HXY}}_z$ by our GSPRG analysis. 
 We consider the case $D=-0.5$ for $J_z \ge 0$. 
 For this purpose, we examine the finite-size exponent
 $\eta_\alpha(\tilde{N},D=-0.5,J_z)$, and estimate 
the critical-ordered boundary point 
$\eta_{\alpha}^{(1)}(N', D=-0.5, J_z)$ 
given by Eq. (\ref{cd-boundary1}) or Eq. (\ref{cd-boundary2}). 
Our results are depicted in Fig. \ref{fig.10}. 
 We find that
 the critical-ordered boundary given by Eq. (\ref{cd-boundary2}) appears 
 when $J_z$ is 0.1, 0.15, 0.18, and 0.2, but it does not appear when 
 $J_z$ is 0. 
Concerning with $x$-component of the string order, 
we find the boundaries at $(J_z=0.18, N' = 29.6)_x$, 
$(J_z=0.15, N' = 31.0)_x$, and $(J_z=0.1, N' = 34.6)_x$. 
Concerning with $z$-component of the string order, 
on the other hand, we have 
$(J_z=0.18, N' = 32.6)_z$, $(J_z=0.15, N' = 35.2)_z$, and 
$(J_z=0.1, N' = 41.3)_z$ as the boundaries. 
In the cases of both of the components, 
one can observe that $N'$ grows when $J_z$ approaches $J_z = 0$. 
These phenomena lead to our result that $J^{\textrm{c,HXY}}_z$ 
is between $J_z = 0$ and $J_z = 0.1$. 
For estimating the transition point more 
 accurately, the critical-ordered boundary point is extrapolated to the 
 limit $N^{\prime}\rightarrow\infty$.
The results are depicted in Fig. \ref{fig.11-5}. 
Since the leading dependence of $J^{\textrm{c, HXY}}_z$ 
on $1/N^{\prime}$ is unknown, 
we here choose the power $1/N^{\prime}$ so that 
the dependence is almost linear. 
We can successfully determine an appropriate value of the power 
for each $\alpha=x$ and $\alpha=z$, 
although the $\alpha=x$ and $\alpha=z$ values differ
from each other. 
A linear extrapolation gives 
$J^{\textrm{c,HXY}}_z(D=-0.5) = 0.00 \pm 0.10$ 
from the transverse component 
and 
$J^{\textrm{c,HXY}}_z(D=-0.5) = 0.01 \pm 0.08$ 
from the longitudinal component. 
Here we determine the error 
as being the difference between the values obtained by the extrapolation and the finite-size critical point $J^{\textrm{c, HXY}}_z$ 
for maximum $N^{\prime}$. 
Both results suggest $J^{\textrm{c,HXY}}_z(D=-0.5) \sim 0$ 
irrespective of the direction of the string order parameter,  
which is consistent with a previous report \cite{Hida:PRB67}. 

Next, we examine what type of transition this is. 
Our finite-size exponents in Fig. \ref{fig.10} 
at $J_z=0$ indicate $\eta_x(D=-0.5, J_z=0) \sim 0.25$ 
and $\eta_z(D=-0.5, J_z=0) \sim 1.0$. 
These values agree well with the exponents 
of the BKT transition $\eta_x = 1/4$ and $\eta_z = 1$.
Our results are also consistent 
with many previous works \cite{Alcaraz:PRB46-2, 
Alcaraz:PRB46, Nomura:JPA28, Hida:PRB67}. 
Therefore, our GSPRG analysis applied 
to the string correlation functions 
is useful in capturing BKT transitions. 

\section{\label{sec:level5}summary and remarks}
We have investigated critical behavior 
near the boundary of the Haldane phase 
in the ground state of an anisotropic $S=1$ chain 
from the viewpoint of string correlation functions 
estimated precisely by standard finite-size DMRG 
under the open boundary condition. 
We have developed the ground-state phenomenological-renormalization-group 
analysis and used it to analyze the correlation functions. 
This analysis provides us with 
the transition point of the boundary of the Haldane phase 
and the critical exponents at and near the transition point. 
Our estimates for these quantities agree with 
those previously obtained from analysis 
of the energy-level structure. 
\begin{figure}[htbp]
	\begin{center}
	\includegraphics[width=7.5cm]{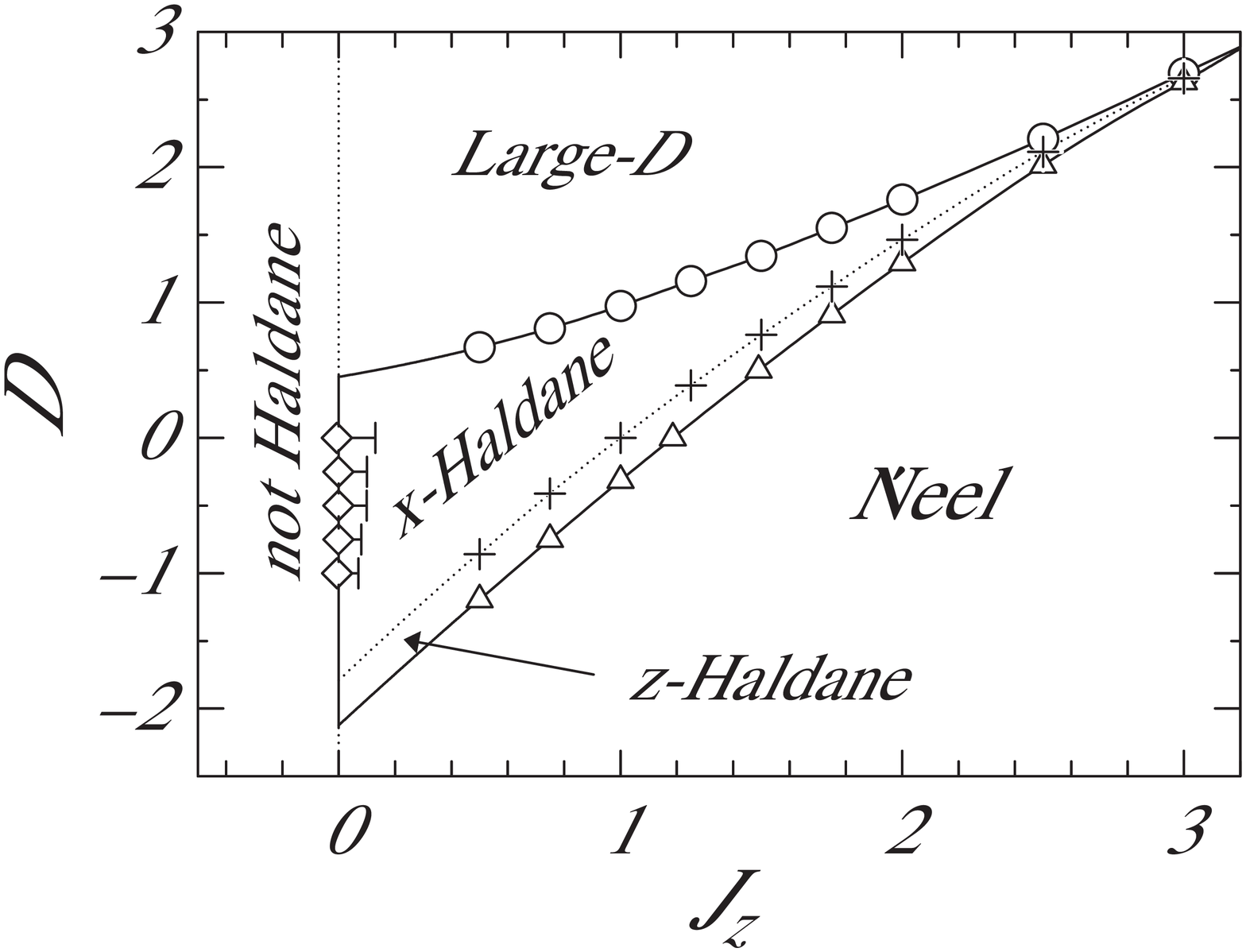}
	\end{center}
	\caption{
Phase diagram for the Hamiltonian (\ref{eq.1}). 
The Haldane--Large-$D$, 
Haldane--N$\acute{\textrm{e}}$el, and Haldane--$XY$ transition 
points are denoted 
by $\bigcirc, \triangle, \Diamond$, respectively. 
All the lines are guides for the eyes. 
x-Haldane and z-Haldane represent 
the $O^x_{str}>O^z_{str}$ region and 
$O^x_{\textrm{str}}<O^z_{\textrm{str}}$ regions, respectively. 
	}
	\label{fig.12}
\end{figure}

We summarize the transition points 
as a ground-state phase diagram in Fig. \ref{fig.12}. 
This figure presents 
the phase boundary of the Haldane--Large-$D$, 
Haldane--N$\acute{\textrm{e}}$el, and Haldane--$XY$ transitions. 
Note additionally that 
the dominant order parameter is $O^x_{\textrm{str}}$ 
in most of the Haldane phase. 

A feature of our approach, GSPRG analysis, 
is that only common quantities under the same condition 
are treated in a unified manner 
irrespective of the type of phase transition. 
Although we have employed the DMRG method in this paper 
to calculate the order parameters, 
we are not limited 
to the DMRG method 
if we can obtain precise estimates of the order parameters.  
The string order parameters of the Haldane phase 
in the $S=1$ chain are examples. 
Other multi-point correlation functions 
for finite-size clusters may also be applicable. 
If we precisely calculate an appropriate ground-state 
quantity that plays the role of an order parameter, 
the framework of the analysis would be widely applicable 
for capturing ground-state critical behavior 
irrespective of the method of calculation and 
the kind of order parameter. 
We hope that the procedure presented in this paper contributes 
to future studies of quantum phase transitions. 

\begin{acknowledgments}
We wish to thank Prof.~K.~Hida, Dr.~K.~Okamoto, Dr.~T.~Sakai, and
Dr.~S.~Todo for fruitful discussions.
We are grateful to Prof.~T.~Nishino, Dr.~K.~Okunishi,
and Dr.~K.~Ueda for comments on the DMRG calculations.
This work was partly supported by Grants-in-Aid
from the Ministry of Education, Culture, Sports,
Science and Technology (Nos. 19310094, 17064006, 20340096),
the 21st COE Program supported by the Japan
Society for Promotion of Science, and
Next Generation Integrated Nanoscience Simulation Software.
A part of the computations was performed using the facilities
of the Information Initiative Center, Hokkaido University
and the Supercomputer Center,
Institute for Solid State Physics, University of Tokyo.
\end{acknowledgments}


\begin{thebibliography}{99}
\bibitem{comment_PBC} 
An estimation of string correlation functions 
with periodic boundary condition by the DMRG method 
has been reported \cite{Boschi:EPJB41}. 
However, the result for correlation functions 
is translationally non-invariant 
in spite of the fact 
that the system is translationally invariant.  
Under such circumstances, 
it is difficult to capture the phase transition precisely 
if we examine the correlation function of the 
longest distance between $i$ and $j=i+N/2$.
\end{thebibliography}

\end{document}